\documentclass[a4paper,fleqn]{cas-sc}

\usepackage[numbers,sort&compress]{natbib}

\usepackage{amsmath}
\usepackage{graphicx}
\usepackage{subcaption}
\usepackage{caption}
\usepackage{textcomp}
\usepackage{amssymb}
\usepackage{cite}
\usepackage{xcolor}
\usepackage{ulem}
\usepackage{siunitx}
\usepackage{pdflscape}

\DeclareMathOperator\sign{sign}
\DeclareMathOperator\artanh{artanh}

\DeclareMathOperator\Sc{Sc}

\def\tsc#1{\csdef{#1}{\textsc{\lowercase{#1}}\xspace}}
\tsc{WGM}
\tsc{QE}
\tsc{EP}
\tsc{PMS}
\tsc{BEC}
\tsc{DE}

\begin{document}
\let\WriteBookmarks\relax
\def\floatpagepagefraction{1}
\def\textpagefraction{.001}
\shorttitle{Full Analytical Solution for the Magnetic Field of Uniformly Magnetized Cylinder Tiles}
\shortauthors{F. Slanovc et~al.}

\title [mode = title]{Full analytical solution for the magnetic field of uniformly magnetized cylinder tiles}                      

\author[1]{F. Slanovc}[orcid=0000-0002-2335-0076]
\cormark[1]
\ead{florian.slanovc@univie.ac.at}

\address[1]{University of Vienna, Boltzmanngasse 5, 1090 Vienna, Austria}

\author[2]{M. Ortner}

\address[2]{Silicon Austria Labs, Europastra\ss e 12, 9524 Villach, Austria}

\author[2]{M. Moridi}

\author[1]{C. Abert}

\author[1]{D. Suess}

\credit{Data curation, Writing - Original draft preparation}

\cortext[cor1]{Corresponding author}

\begin{abstract}
We present an analytical solution for the magnetic field of a homogeneously magnetized cylinder tile and by extension solutions for full cylinders, rings, cylinder sectors and ring segments. The derivation is done by direct integration in the magnetic surface charge picture. Results are closed-form expressions and elliptic integrals. All special cases are treated individually, which enables the field computation for all possible position arguments ${\bf r}\in\mathbb{R}^3$. An implementation is provided in Python together with a performance analysis. The implementation is tested against numerical solutions and applied to compute the magnetic field in a discrete Halbach cylinder. 
\end{abstract}


\maketitle

\section{Introduction}\label{sec::intro}
Permanent magnets are widely used in numerous technical applications \citep{strnat1990, fastenau1996, stekly2000, coey2002, mccallum2014, hirosawa2017, coey2020, b52}. Among all conceivable geometries, simple magnet shapes such as cuboids or cylinders are especially popular and widespread as they are mass-produced and available at short notice in various dimensions, magnetization and materials \citep{b31, b32}. When the magnet geometry is simple and the magnetization is homogeneous, it is often possible to derive analytical expressions for the field generated by such magnets \citep{b25,b26,b27,b33,b51,b35,b36,b37,b38,b39}. These expressions provide excellent approximations to the real fields when modern, high-grade magnetic materials, like SmCo, NdFeB or ferrites with susceptibilities $\chi<0.1$ are involved \citep{b52,b30}.

The main advantage over common numerical methods such as finite element (FE) approaches or direct numerical integration is the fast computation times of the order of microseconds \citep{b30,b77}, which enables highly efficient multivariate parameter space analysis and solving global optimization problems for permanent magnet arrangements \citep{b46, b47, b49}.

The analytical formulas for cuboid magnets \citep{b33,b52,b60,engelherbert2005} are particularly useful since cuboids can be trivially subdivided into smaller cuboid cells to which the formulas are again applicable.
This allows modeling of the material response with the magnetostatic method of moments \citep{b41}, a common approach of computing dipole interactions in micromagnetic software \citep{b53,b54,b55,b56,b57,b58,b76}.
In contrast, the analytical expression of a cylinder \citep{b51, b35, b36, b37, b38, b39} is not sufficient for such calculations, as the cylindrical volume element is a cylinder tile, i.e., a cylindrical ring segment. In addition, many geometric shapes used in magnetic position systems, such as the cylindrical quadrupole \citep{b59,b61} or rotary encoder wheels \citep{b62, iso}, can be constructed with such tiles, motivating us to find analytical expressions for this geometry.

Finding an analytical expression for the homogeneously magnetized cylinder tile was attempted in many previous works \citep{b9,b10,b11,b12,b13,b14}, but to date all solutions contain integral expressions which can only be solved numerically.
In this work we present for the first time a complete analytical solution for the magnetic field of a homogeneously magnetized cylinder tile. All expressions contain only basic arithmetic operations, trigonometric functions and special functions like elliptic integrals, which can be computed very efficiently \citep{b6,b7,b63,b64,b65,b66,b67,b68,b69,b70}.
In previous works, only the most general case is treated, while indeterminate forms, that typically appear at surface extensions, are neglected. We provide expressions for all such cases, which allows us to compute the field at any point in space. In addition, we include the limiting cases with inner radius $r_1 = 0$ (cylinder sector), constricting sector angles $\varphi_1 = 0$, $\varphi_2 = 2\pi$ (cylinder ring) and the full cylinder as the combination of the two cases.



The structure of the paper is as follows: In Sec.~\ref{sec::method} we establish the notation and describe the surface charge method, which enables us to express the total magnetic field as the sum of various surface integrals.
The first and second integration steps are shown in Sec.~\ref{sec::1stint} and Sec.~\ref{sec::2ndint}, respectively.
In Sec.~\ref{sec::results} we validate our results by comparison with solutions from numerical integration and finite element method. A discrete Halbach cylinder application example is demonstrated in Sec.~\ref{sec::halbach}.
We close with a brief discussion of the results and outlook in Sec.~\ref{sec::outlook}.

\section{Method}\label{sec::method}
\begin{figure}
    \centering
    \begin{subfigure}[b]{0.49\textwidth}
        \centering
        \includegraphics[width=1.0\textwidth]{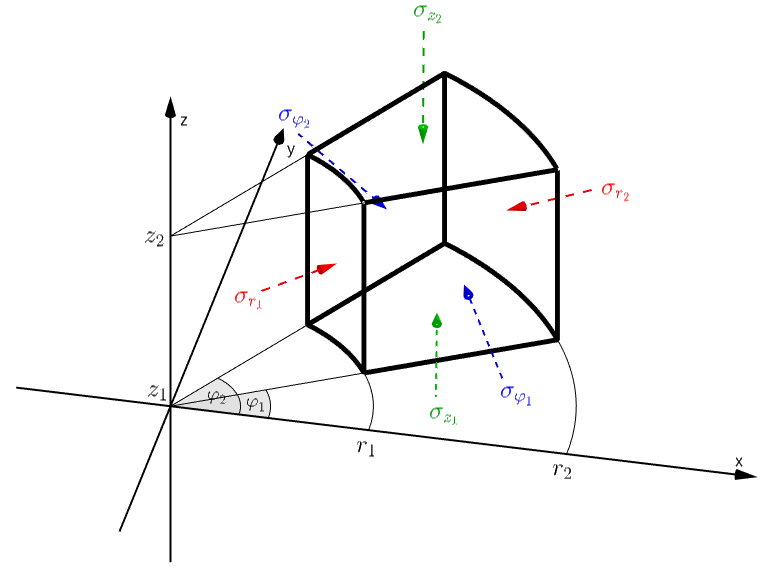}
        \caption{}
        \label{fig::tile}
    \end{subfigure}
    \hfill
    \begin{subfigure}[b]{0.49\textwidth}
        \centering
        \includegraphics[width=1.0\textwidth]{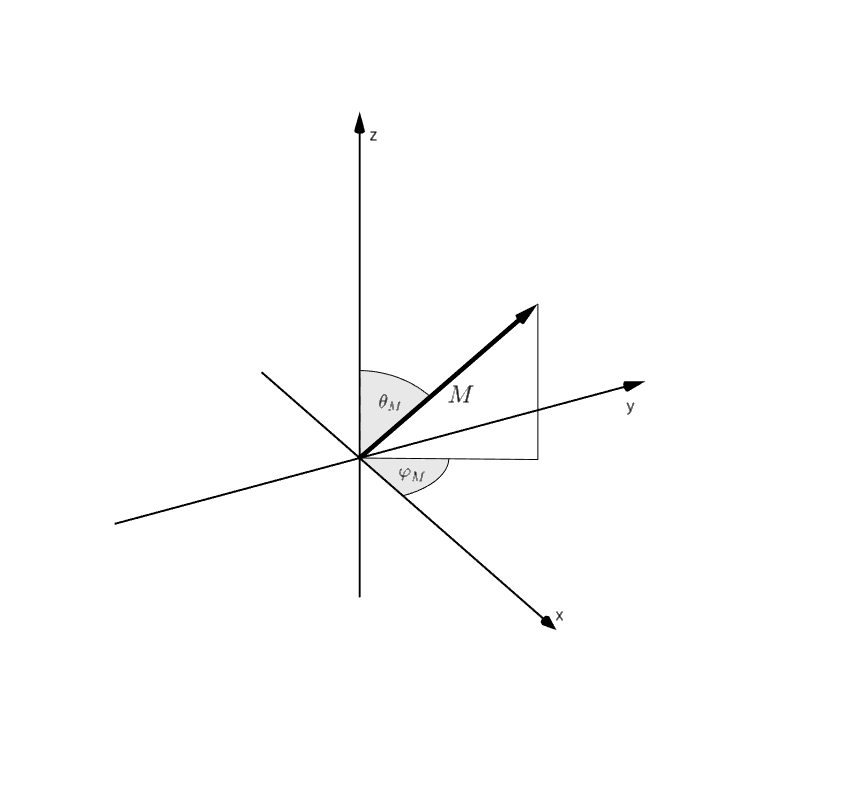}
        \caption{}
        \label{fig::J}
    \end{subfigure}
    \caption{(a) Dimensions of the cylinder tile in cylinder coordinates and the resulting magnetic surface charges $\sigma$. We assume $0\leq r_1 < r_2$, $\varphi_1<\varphi_2$ and $z_1<z_2$. (b) Magnetization vector ${\bf M}$ described by the three coordinates $M, \varphi_M$ and $\theta_M$.}
\end{figure}

 It is convenient to describe the problem in cylindrical coordinates $(r,\varphi,z) \in [0,\infty)\times [0,2\pi) \times (-\infty,\infty)$. The cylinder tile is defined through six boundaries at $r_1<r_2$, $\varphi_1<\varphi_2$, and $z_1<z_2$, see Fig.~\ref{fig::tile}. We define the homogeneous magnetization vector $\bf{M}$, through its magnitude $M$, azimuth angle $\varphi_M$ and polar angle $\theta_M$, see Fig.~\ref{fig::J}.

In Cartesian coordinates the magnetization is then given by
\begin{align}\label{eq::J}
    {\bf M} = M \left(\cos{\varphi_M} \sin{\theta_M} {\bf e}_x + \sin{\varphi_M} \sin{\theta_M} {\bf e}_y + \cos{\theta_M} {\bf e}_z\right).
\end{align}
 
%


%

\subsection{Field calculation with magnetostatic potential}
In \citep[Sec.~5.9 C]{b3}, the scalar magnetic potential $\Phi_M$ is derived for volumes $V$ with homogeneous magnetization ${\bf M}$ in the absence of electric currents,
%
\begin{align}
    \Phi_M({\bf r}) = \frac{1}{4 \pi} \oint_{\partial V} \frac{\sigma({\bf r}')} {\left|{\bf r}-{\bf r}'\right|}\ \text{d}A',
\end{align}
with the magnetic surface charge $\sigma$ given by the inner product of the magnetization vector with the surface normal vector ${\bf n}$,
\begin{align}\label{eq::sigma}
\sigma({\bf r}') = {\bf M}({\bf r}') \cdot {\bf n}({\bf r}').
\end{align}
The magnetic field ${\bf H}$ is then given by
\begin{align}\label{eq::H_cart}
    {\bf H}({\bf r}) = -\boldsymbol{\nabla}\Phi_M({\bf r}) = \frac{1}{4 \pi} \oint_{\partial V} \frac{\sigma({\bf r}') ({\bf r}-{\bf r}')} {\left|{\bf r}-{\bf r}'\right|^3}\ \text{d}A'.
\end{align}

\subsection{Magnetic surface-charge density}
The tile boundary $\partial V$ includes three types of surfaces. Each type is denoted by its normal vector, see Fig.~\ref{fig::tile}. We calculate the magnetic surface charge density $\sigma$ according to Eq.~\eqref{eq::sigma} in cylindrical coordinates for each surface type.
%
\begin{align}
    \begin{split}
        \sigma_{r_i}(\varphi') &= (-1)^{i} {\bf M} \cdot \left(\cos{\varphi'} {\bf e}_x + \sin{\varphi'} {\bf e}_y \right) \\ &= (-1)^{i} M \sin{\theta_M} \left( \cos{\varphi_M}  \cos{\varphi'} + \sin{\varphi_M}  \sin{\varphi'}\right) \\ &= (-1)^{i} M \sin{\theta_M} \cos(\varphi_M-\varphi')
    \end{split}\\
    \begin{split}
        \sigma_{\varphi_j} &= (-1)^{j} {\bf M}\cdot \left(-\sin{\varphi_j} {\bf e}_x + \cos{\varphi_j} {\bf e}_y \right) \\ &= (-1)^{j} M \sin{\theta_M} \left( -\cos{\varphi_M}  \sin{\varphi_j} + \sin{\varphi_M}  \cos{\varphi_j}\right) \\ &= (-1)^{j} M \sin{\theta_M} \sin(\varphi_M-\varphi_j)
    \end{split}\\
    \begin{split}
        \sigma_{z_k} &= (-1)^{k} {\bf M}\cdot  {\bf e}_z \\
        &= (-1)^{k} M\cos{\theta_M}
    \end{split}
\end{align}

It will be useful for our derivation to omit magnetization and powers of $(-1)$ and define the dashed (sign reduced) surface charge densities by
%
\begin{align}\label{eq::reduced}
    \sigma_{r_i}'(\varphi') & = \sin{\theta_M} \cos(\varphi_M-\varphi'),\\
    \sigma_{\varphi_j}' & = \sin{\theta_M} \sin(\varphi_M-\varphi_j),\\
    \sigma_{z_k}' & = \cos{\theta_M}.
\end{align}

\subsection{Surface integrals}
To express the the surface integrals \eqref{eq::H_cart} in cylindrical coordinates, we use the usual transformations,
\begin{align}
    x &= r\cos{\varphi},\label{eq::trans1} \\
    y &= r\sin{\varphi},\label{eq::trans2}\\
    {\bf e}_r &=  \cos\varphi\ {\bf e}_x + \sin\varphi\ {\bf e}_y, \label{eq::trans3} \\
    {\bf e}_\varphi &=  -\sin\varphi\ {\bf e}_x + \cos\varphi\ {\bf e}_y. \label{eq::trans4}
\end{align}
To simplify the notation, we do not explicitly write the dependency of the coordinates $r,\varphi$ and $z$, although almost every term depends on them. With the abbreviation
\begin{align}
    \xi(r',\varphi',z'):=\left|{\bf r}-{\bf r}'\right| = \sqrt{r^2+r'^2-2rr'\cos(\varphi-\varphi')+(z-z')^2},
\end{align}
every field component can be expressed as a sum of six terms, each originating from one tile surface, as

\begin{align}\label{eq::Hr}
\begin{split}
     H_r = \frac{M}{4 \pi} \Bigg(& \sum_{i=1}^2 (-1)^i \int_{\varphi_1}^{\varphi_2}\int_{z_1}^{z_2} \underbrace{\frac{\sigma_{r_i}'(\varphi') (r - r_i\cos(\varphi-\varphi'))} {\xi(r_i,\varphi',z')^3}r_i}_{=:H_{r,r_i}(\varphi',z')} \ \text{d}z'\text{d}\varphi' + \\
    &\sum_{j=1}^2 (-1)^j \int_{r_1}^{r_2}\int_{z_1}^{z_2} \underbrace{\frac{\sigma_{\varphi_j}' (r - r'\cos(\varphi-\varphi_j))} {\xi(r',\varphi_j,z')^3}}_{=:H_{r,\varphi_j}(r',z')} \ \text{d}z'\text{d}r' + \\
   & \sum_{k=1}^2 (-1)^k \int_{\varphi_1}^{\varphi_2}\int_{r_1}^{r_2} \underbrace{\frac{\sigma_{z_k}' (r - r'\cos(\varphi-\varphi'))} {\xi(r',\varphi',z_k)^3}r'}_{=:H_{r,z_k}(r',\varphi')}\ \text{d}r'\text{d}\varphi'\Bigg) 
   \end{split}
\end{align}
\begin{align}\label{eq::Hphi}
\begin{split}
 H_\varphi = \frac{M}{4 \pi} \Bigg(& \sum_{i=1}^2 (-1)^i \int_{\varphi_1}^{\varphi_2}\int_{z_1}^{z_2} \underbrace{ \frac{\sigma_{r_i}(\varphi') r_i^2\sin(\varphi-\varphi')} {\xi(r_i,\varphi',z')^3}}_{=:H_{\varphi,r_i}(\varphi',z')} \ \text{d}z'\text{d}\varphi' + \\
    &\sum_{j=1}^2 (-1)^j \int_{r_1}^{r_2}\int_{z_1}^{z_2} \underbrace{\frac{\sigma_{\varphi_j} r'\sin(\varphi-\varphi_j)} {\xi(r',\varphi_j,z')^3}}_{=:H_{\varphi,\varphi_j}(r',z')} \ \text{d}z'\text{d}r' + \\
   & \sum_{k=1}^2 (-1)^k \int_{\varphi_1}^{\varphi_2}\int_{r_1}^{r_2} \underbrace{\frac{\sigma_{z_k} r'^2\sin(\varphi-\varphi')} {\xi(r',\varphi',z_k)^3}}_{=:H_{\varphi,z_k}(r',\varphi')}\ \text{d}r'\text{d}\varphi'\Bigg)
   \end{split}
\end{align}
\begin{align}\label{eq::Hz}
\begin{split}
    H_z = \frac{M}{4 \pi} \Bigg(& \sum_{i=1}^2 (-1)^i \int_{\varphi_1}^{\varphi_2}\int_{z_1}^{z_2} \underbrace{\frac{\sigma_{r_i}(\varphi') (z-z')} {\xi(r_i,\varphi',z')^3}r_i}_{=:H_{z,r_i}(\varphi',z')} \ \text{d}z'\text{d}\varphi' + \\
    &\sum_{j=1}^2 (-1)^j \int_{r_1}^{r_2}\int_{z_1}^{z_2} \underbrace{\frac{\sigma_{\varphi_j} (z-z')} {\xi(r',\varphi_j,z')^3}}_{=:H_{z,\varphi_j}(r',z')} \ \text{d}z'\text{d}r' + \\
   & \sum_{k=1}^2 (-1)^k \int_{\varphi_1}^{\varphi_2}\int_{r_1}^{r_2} \underbrace{\frac{\sigma_{z_k} (z-z_k)} {\xi(r',\varphi',z_k)^3}r'}_{=:H_{z,z_k}(r',\varphi')}\ \text{d}r'\text{d}\varphi'\Bigg) 
   \end{split}
\end{align}

In the following sections, we present the analytical solution of the above nine surface integrals.
%
%
In our derivations below, we denote the solution to the above integrals using upper indices, e.g.
\begin{align}
    \int_{z_1}^{z_2} H_{r,r_i}(\varphi',z') \text{d}z' =  H_{r,r_i}^{z_2}(\varphi')-H_{r,r_i}^{z_1}(\varphi')
\end{align}
which allows us to express the final solution in a very compact manner,
\begin{align} \label{eq::totalfield}
    H_\alpha  = \frac{M}{4 \pi} \sum_{i,j,k=1}^2 (-1)^{i+j+k} \big(H_{\alpha,r_i}^{\varphi_j,z_k} + H_{\alpha,\varphi_j}^{r_i,z_k} + H_{\alpha,z_k}^{r_i,\varphi_j} \big)
\end{align}
for $\alpha\in\{r,\varphi,z\}$.

\section{First integration}\label{sec::1stint}
The first integration can be easily achieved with modern computer algebra systems and was demonstrated in previous works \citep{b10}. To simplify the notation, we introduce the following short forms:
\begin{align}
    \overline{r}_i&:=r-r_i\\
    \overline{r}'&:=r-r'\\
    \overline{\varphi}_j&:=\varphi-\varphi_j\\
    \overline{\varphi}'&:=\varphi-\varphi'\\
    \overline{\varphi}_M&:=\varphi_M-\varphi\\
    \overline{\varphi}_M'&:=\varphi_M-\varphi'\\ 
    \overline{\varphi}_{Mj}&:=\varphi_M-\varphi_j\\ 
    \overline{z}_k&:=z-z_k\\
    \overline{z}'&:=z-z'
\end{align}
The first integration then yields:
 \begin{landscape}
\begin{align}
    H_{r,r_i}^{z_k}(\varphi') &= \left[\int H_{r,r_i}(\varphi',z')\ \text{d}z'\right]_{z'=z_k} = \left\{ \begin{array}{ll}
       -\sin{\theta_M}r_i\overline{z}_k\frac{\cos\overline{\varphi}_M'(r-r_i\cos\overline{\varphi}')}{(r^2+r_i^2-2rr_i\cos\overline{\varphi}')\xi(r_i,\varphi',z_k)} & \text{ for } r\neq r_i \text{ and } r>0 \\
        -\frac{\sin{\theta_M}\overline{z}_k}{2}\frac{\cos\overline{\varphi}_M'}{\sqrt{2r^2(1-\cos\overline{\varphi}')+\overline{z}_k^2}} & \text{ for } r = r_i \text{ and } r>0 \\
        \frac{\sin{\theta_M}\overline{z}_k}{\sqrt{r_i^2+\overline{z}_k^2}}\cos\overline{\varphi}_M'\cos\overline{\varphi}' &  \text{ for }  r=0\text{ and } r_i>0\\
        0 & \text{ for }  r_i=0
    \end{array} \right. \label{eq::1stintstart} \\
    H_{r,\varphi_j}^{r_i}(z') &= \left[\int H_{r,\varphi_j}(r',z')\ \text{d}r'\right]_{r'=r_i} = \left\{ \begin{array}{ll} 2\sin{\theta_M}\sin\overline{\varphi}_{Mj}\frac{\overline{z}'^2\cos \overline{\varphi}_j+r r_i\sin^2\overline{\varphi}_j}{(r^2+2\overline{z}'^2-r^2\cos(2\overline{\varphi}_j))\xi(r_i,\varphi_j,z')} & \text{ for } \overline{\varphi}_j\not\in\pi\mathbb Z   \text{ and } r>0 \\
    \sin{\theta_M}\sin\overline{\varphi}_{Mj}\frac{1}{\sqrt{\overline{r}_i^2+\overline{z}'^2}} & \text{ for } \overline{\varphi}_j\in 2\pi\mathbb Z
     \text{ and } r>0 \\
    -\sin{\theta_M}\sin\overline{\varphi}_{Mj}\frac{1}{\sqrt{(r+r_i)^2+\overline{z}'^2}} & \text{ for } \overline{\varphi}_j\in 2\pi\mathbb Z+\pi
     \text{ and } r>0 \\
    \sin{\theta_M}\sin\overline{\varphi}_{Mj}\frac{\cos\overline{\varphi}_j}{\sqrt{r_i^2+\overline{z}'^2}} & \text{ for }  r=0 
    \end{array} \right. \\
    H_{r,z_k}^{r_i}(\varphi') &= \left[\int H_{r,z_k}(r',\varphi')\ \text{d}r'\right]_{r'=r_i} = 
    \left\{ \begin{array}{ll} 
    \cos\theta_M ( \frac{r_i(r^2+2\overline{z}_k^2)\cos\overline{\varphi}' - r(r^2+\overline{z}_k^2-(r^2+\overline{z}_k^2)\cos(2\overline{\varphi}') - r r_i \cos(3\overline{\varphi}'))}{(r^2+2\overline{z}_k^2-r^2\cos(2\overline{\varphi}'))\xi(r_i,\varphi',z_k)} & \\
    - \cos\overline{\varphi}' \log(r_i-r\cos\overline{\varphi}'+\xi(r_i,\varphi',z_k)) ) & \text{ for } z\neq z_k  \text{ and } r>0  \\
    \cos\theta_M ( \frac{2r_i\cos\overline{\varphi}'-r}{\sqrt{r^2+r_i^2-2r r_i\cos\overline{\varphi}'}} & \\
    - \cos\overline{\varphi}' \log(r_i-r\cos\overline{\varphi}'+\sqrt{r^2+r_i^2-2r r_i\cos\overline{\varphi}'})) & \text{ for } z= z_k   \text{ and } r>0 \\
    \cos\theta_M ( \frac{r_i}{\sqrt{r_i^2+\overline{z}_k^2}} - \artanh(\frac{r_i}{\sqrt{r_i^2+\overline{z}_k^2}}) )\cos\overline{\varphi}' & \text{ for } z\neq z_k  \text{ and } r=0 \\
    -\cos\theta_M \log r_i \cos\overline{\varphi}' & \text{ for } z = z_k  \text{ and } r=0
    \end{array} \right. \label{eq::rz} \\
    H_{\varphi,r_i}^{z_k}(\varphi') &= \left[\int H_{\varphi,r_i}(\varphi',z')\ \text{d}z'\right]_{z'=z_k} = \left\{ \begin{array}{ll}
       -\sin{\theta_M}r_i^2\overline{z}_k\frac{\cos\overline{\varphi}_M'\sin\overline{\varphi}'}{(r^2+r_i^2-2rr_i\cos\overline{\varphi}')\xi(r_i,\varphi',z_k)} & \text{ for } r\neq r_i \text{ and } r>0  \\
       -\frac{\sin{\theta_M}}{2}\frac{\cos\overline{\varphi}_M'\sin\overline{\varphi}'}{(1-\cos\overline{\varphi}')} \frac{\overline{z}_k}{\sqrt{2r^2(1-\cos\overline{\varphi}')+\overline{z}_k^2}}  & \text{ for } r = r_i \text{ and } r>0 \text{ and } \overline{\varphi}_j \not\in 2\pi\mathbb Z \text{ for } j=1 \text{ and } j=2 \\
        -\frac{\sin{\theta_M}}{2}\frac{\cos\overline{\varphi}_M'\sin\overline{\varphi}'}{(1-\cos\overline{\varphi}')}\left( \frac{\overline{z}_k}{\sqrt{2r^2(1-\cos\overline{\varphi}')+\overline{z}_k^2}} - \sign\overline{z}_k \right) & \text{ for } r = r_i \text{ and } r>0 \text{ and } \overline{\varphi}_j \in 2\pi\mathbb Z \text{ for } j=1 \text{ or } j=2 \\
         -\frac{\sin{\theta_M}\overline{z}_k}{\sqrt{r_i^2+\overline{z}_k^2}}\cos\overline{\varphi}_M'\sin\overline{\varphi}' & \text{ for } r=0 \text{ and } r_i>0 \\
         0 & \text{ for } r_i=0
    \end{array} \right. \label{eq::phir} \\
        H_{\varphi,\varphi_j}^{z_k}(r') &= \left[\int H_{\varphi,\varphi_j}(r',z')\ \text{d}z'\right]_{z'=z_k} = \left\{ \begin{array}{ll}
       -\sin{\theta_M}\sin\overline{\varphi}_{Mj}\sin\overline{\varphi}_j\overline{z}_k\frac{r'}{(r^2+r'^2-2rr'\cos\overline{\varphi}_j)\xi(r',\varphi_j,z_k)} & \text{ for } \overline{\varphi}_j\not\in \pi\mathbb Z  \\
         0 & \text{ for } \overline{\varphi}_j\in \pi\mathbb Z
    \end{array} \right.   \\
     H_{\varphi,z_k}^{\varphi_j}(r') &= \left[\int H_{\varphi,z_k}(r',\varphi')\ \text{d}\varphi'\right]_{\varphi'=\varphi_j} = 
     \left\{ \begin{array}{ll}
     \frac{\cos{\theta_M}}{r}\frac{r'}{\xi(r',\varphi_j,z_k)} & \text{ for } r>0 \\
     \cos{\theta_M}\cos\overline{\varphi}_j\frac{r'^2}{\sqrt{r'^2+\overline{z}_k^2}^3} & \text{ for } r=0
     \end{array} \right. \\ 
      H_{z,r_i}^{z_k}(\varphi') &= \left[\int H_{z,r_i}(\varphi',z')\ \text{d}z'\right]_{z'=z_k} = \sin{\theta_M} r_i \frac{\cos\overline{\varphi}_M'}{\xi(r_i,\varphi',z_k)} \\
      H_{z,\varphi_j}^{z_k}(r') &= \left[\int H_{z,\varphi_j}(r',z')\ \text{d}z'\right]_{z'=z_k} =  \sin{\theta_M}\sin\overline{\varphi}_{Mj}\frac{1}{\xi(r',\varphi_j,z_k)}  \\
      H_{z,z_k}^{r_i}(\varphi') &= \left[\int H_{\varphi,z_k}(r',\varphi')\ \text{d}r'\right]_{r'=r_i} = 
     \left\{ \begin{array}{ll}
           -2\cos{\theta_M}\overline{z}_k\frac{r^2+\overline{z}_k^2-r r_i\cos\overline{\varphi}'}{(r^2+2\overline{z}_k^2-r^2\cos(2\overline{\varphi}'))\xi(r_i,\varphi',z_k)}  & \text{ for } z\neq z_k \\
     0 & \text{ for } z = z_k
     \end{array} \right. \label{eq::1stintend} \\ 
\end{align}
 \end{landscape}

The first integration leads to several singularities which must be treated with care within the second integration step. In Appendix~\ref{app::singularies} we discuss these problems in detail. 

\section{Second integration}\label{sec::2ndint}
%
With the second integration step, the number of special cases to be distinguished increases massively.
%
To identify the special cases outlined in Tab.~\ref{tab::cases}, a three-digit index $\mathcal{I} := \ell mn$ is introduced. For example, the index $\mathcal{I}=213$ denotes the case $z \neq z_k$, $\overline{\varphi}_j \in 2\pi\mathbb Z$, $r>0$, $r_i=0$. The special cases $\mathcal{I} = 111, 114, 121, 131$ are unphysical as they refer to a field evaluation on corners and edges of the tile.
\begin{table}
\begin{center}
\begin{tabular}{ c||c|c|c } 
& $\ell$ & $m$ & $n$ \\
 \hline
 $1$ &  $z = z_k$ & $\overline{\varphi}_j \in 2\pi\mathbb Z$ & $r=r_i=0$ \\ 
 $2$  & $z \neq z_k$ &  $\overline{\varphi}_j \in 2\pi\mathbb Z + \pi$ & $r=0, r_i>0$\\ 
 $3$ & & $\overline{\varphi}_j \not\in\pi\mathbb Z$ & $r>0,r_i=0$  \\ 
 $4$ &  & & $r=r_i>0$  \\ 
 $5$ & & & $r,r_i>0, r\neq r_i$  
\end{tabular}
\end{center}
\caption{All possible values of the three-digits index $\mathcal{I}=\ell mn$ and associated special cases. The general case is $\mathcal{I} = 235$, i.e. $z \neq z_k$, $\overline{\varphi}_j \not\in\pi\mathbb Z$ and $r,r_i>0, r\neq r_i$.}
\label{tab::cases}
\end{table}

For each case we provide a table with all expressions and coefficients that make up the components of \eqref{eq::totalfield}. An individual field component is computed by summing over all functions multiplied by coefficients given in the tables. An example is demonstrated for the case $\mathcal{I}=211$ with Tab.~\ref{tab::examplecase} from which the field components calculate as

\begin{table}
\begin{center}
\begin{tabular}{ c||c|c } 
& $\varphi_j$ & 
$\sign{\overline{z}_k} \log |\overline{z}_k|$ 
\\
 \hline\hline
 $H_{r,\varphi_j}^{r_i,z_k}$ & 
 $0$ & 
 $-\sin{\theta_M} \sin{\overline{\varphi}_M}$ 
 \\ 
 \hline
 $H_{z,z_k}^{r_i,\varphi_j}$ & 
 $-\cos{\theta_M} \sign{\overline{z}_k}$ & 
 $0$
\end{tabular}
\end{center}
\caption{Example table for $\mathcal{I} = 211$. The field components are computed by summation over all functions multiplied by coefficients.}
\label{tab::examplecase}
\end{table}

\begin{align}
H_{r,\varphi_j}^{r_i,z_k} &= 0 \cdot \varphi_j - \sin{\theta_M} \sin{\overline{\varphi}_M} \cdot \sign{\overline{z}_k} \log |\overline{z}_k|, \\
H_{z,z_k}^{r_i,\varphi_j} &= -\cos{\theta_M} \sign{\overline{z}_k} \cdot \varphi_j + 0 \cdot \sign{\overline{z}_k} \log |\overline{z}_k|,
\end{align}
while all other field components not indicated are equal to zero in this case, i.e.
\begin{align}
H_{r,r_i}^{\varphi_j,z_k} = H_{r,z_k}^{r_i,\varphi_j} = H_{\varphi,r_i}^{\varphi_j,z_k} = H_{\varphi,\varphi_j}^{r_i,z_k} = H_{\varphi,z_k}^{r_i,\varphi_j} = H_{z,r_i}^{\varphi_j,z_k} = H_{z,\varphi_j}^{r_i,z_k} = 0.
\end{align}
The tables for all cases can be found in Appendix~\ref{app::tables}.

\section{Numerical verification}\label{sec::results}

\subsection{Implementation and performance}
The analytical formulas are lengthy and tedious to implement numerically.
For public use, we have integrated our fully tested and vectorized code into the upcoming version 4 of the open-source package Magpylib \citep{b30} under the source name \verb CylinderSegment. Although the program flow of the implementation is, at this time, not optimal, we still achieve computation times of the order of few milliseconds (single points) and few tens of microseconds (vectorized evaluation of 2000 points) on mobile CPUs such as Intel i5-8365U, 1.60 GHZ. In comparison, a computation by numerical integration with the precision of Sec.~\ref{sec::quad} takes a few tens of seconds (single points, no vectorization possible), while the performance trimmed finite element code from Sec.~\ref{sec::fem} takes several hours (complete solution) on the same machine.

\subsection{Comparison to numerical integration}
\label{sec::quad}

\begin{figure}
    \centering
    \begin{subfigure}[b]{0.34\textwidth}
        \centering
        \includegraphics[width=1.0\textwidth]{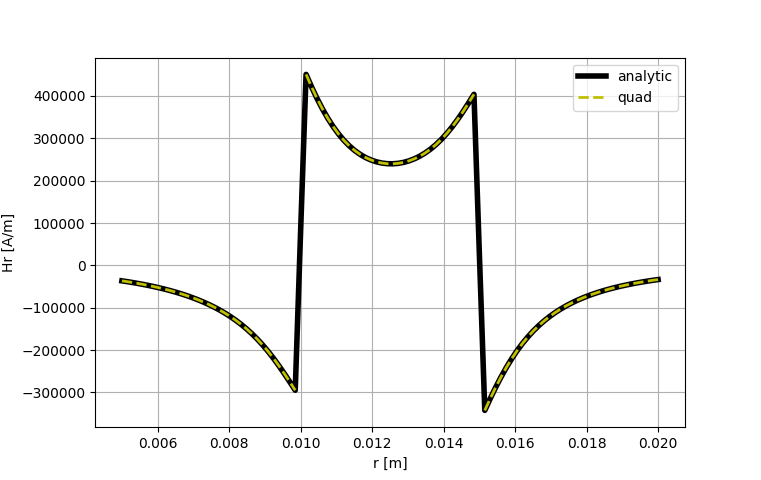}
        \caption{}
        \label{fig::Hr}
    \end{subfigure}~\begin{subfigure}[b]{0.34\textwidth}
        \centering
        \includegraphics[width=1.0\textwidth]{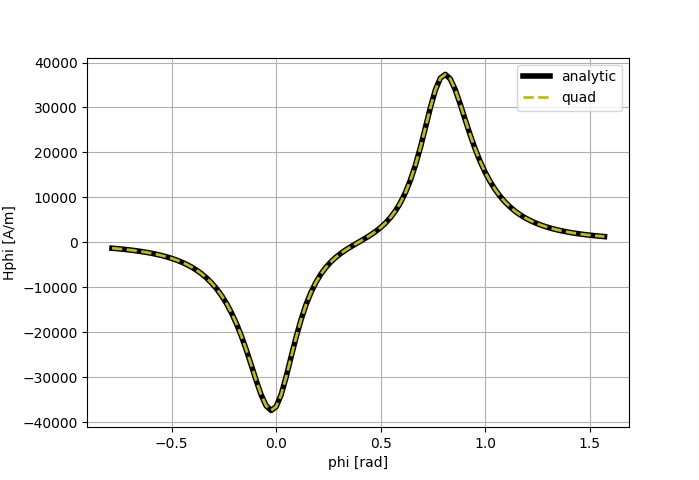}
        \caption{}
        \label{fig::Hphi}
    \end{subfigure}~\begin{subfigure}[b]{0.34\textwidth}
    \centering
        \includegraphics[width=1.0\textwidth]{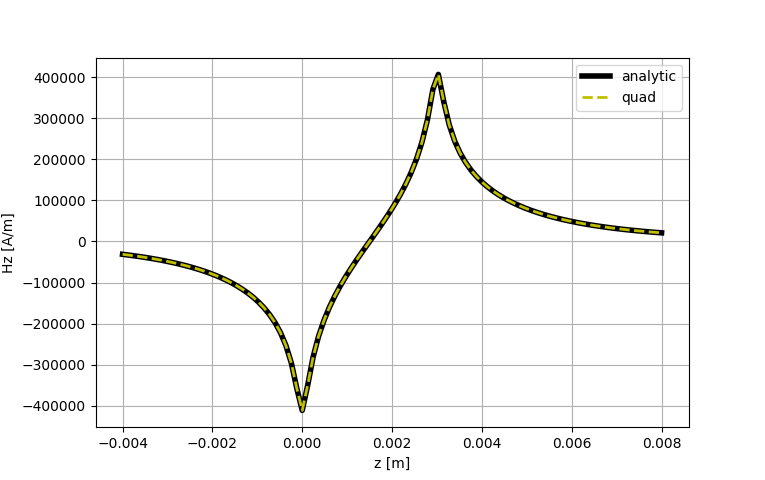}
        \caption{}
        \label{fig::Hz}
    \end{subfigure}
    \caption{Comparison between the derived analytical solution in Sec.~\ref{sec::2ndint} (analytic) and the direct numerical integration of the integrals in Eqs.~\eqref{eq::Hr}-\eqref{eq::Hz} with quadrature (quad). (a) $H_r$ component as function of radial coordinate $r$ at $\varphi = \SI{\pi/8}{\radian}$ and $z = \SI{0.0015}{\m}$ for test geometry 1. (b) $H_\varphi$ component as function of angular coordinate $\varphi$ at $r = \SI{0.022}{\m}$ and $z = \SI{0.001}{\m}$ for test geometry 2 (c) $H_z$ component as function of angular coordinate $z$ at $r = \SI{0.0249}{\m}$ and $\varphi = \SI{\pi/8}{\radian}$ for test geometry 2.}
    \label{fig::numint}
\end{figure}

We demonstrate the correctness of the solution and our implementation by comparing with direct numerical integration via \texttt{scipy.integrate.dblquad} \citep{b71} of the integrals \eqref{eq::Hr}-\eqref{eq::Hz}. The dimensions of the cylinder tiles and field evaluation points are chosen similarly to \citep[Figs.~6-8]{b10} and are listed in Tab.~\ref{tab::geometries} and the results in Fig.~\ref{fig::numint}. All graphs are in perfect agreement. 

\begin{table}
\begin{center}
\begin{tabular}{ c||c|c|c|c|c|c|c|c|c } 
 & $r_1$ [mm] & $r_2$ [mm] & $\varphi_1$ [rad] & $\varphi_2$ [rad] & $z_1$ [mm] & $z_2$ [mm] & $\mu_0 M$ [T] & $\varphi_M$ [rad] & $\theta_M$ [rad]  \\
 \hline
test geometry 1 & $10$ & $15$ & $0$ & $\pi / 4$ & $0$ & $3$ & $1$ &  $9\pi/8$ & $\pi/2$  \\
test geometry 2 & $25$ & $30$ & $0$ & $\pi / 4$ & $0$ & $3$ & $1$ & $9\pi/8$ & $\pi/2$ 
\end{tabular}
\end{center}
\caption{Test geometries for comparison to numerical integration.}
\label{tab::geometries}
\end{table}

\subsection{Comparison to finite elements method}\label{sec::fem}
\begin{figure}
    \centering
    \begin{subfigure}[b]{0.45\textwidth}
        \centering
        \includegraphics[width=0.9\textwidth]{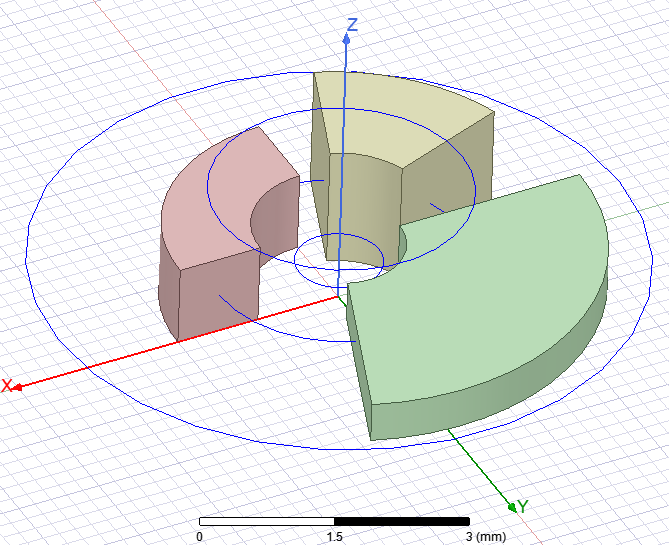}
        \caption{}
        \label{fig::tiles}
    \end{subfigure}
    \hfill
    \begin{subfigure}[b]{0.54\textwidth}
        \includegraphics[width=.99\textwidth]{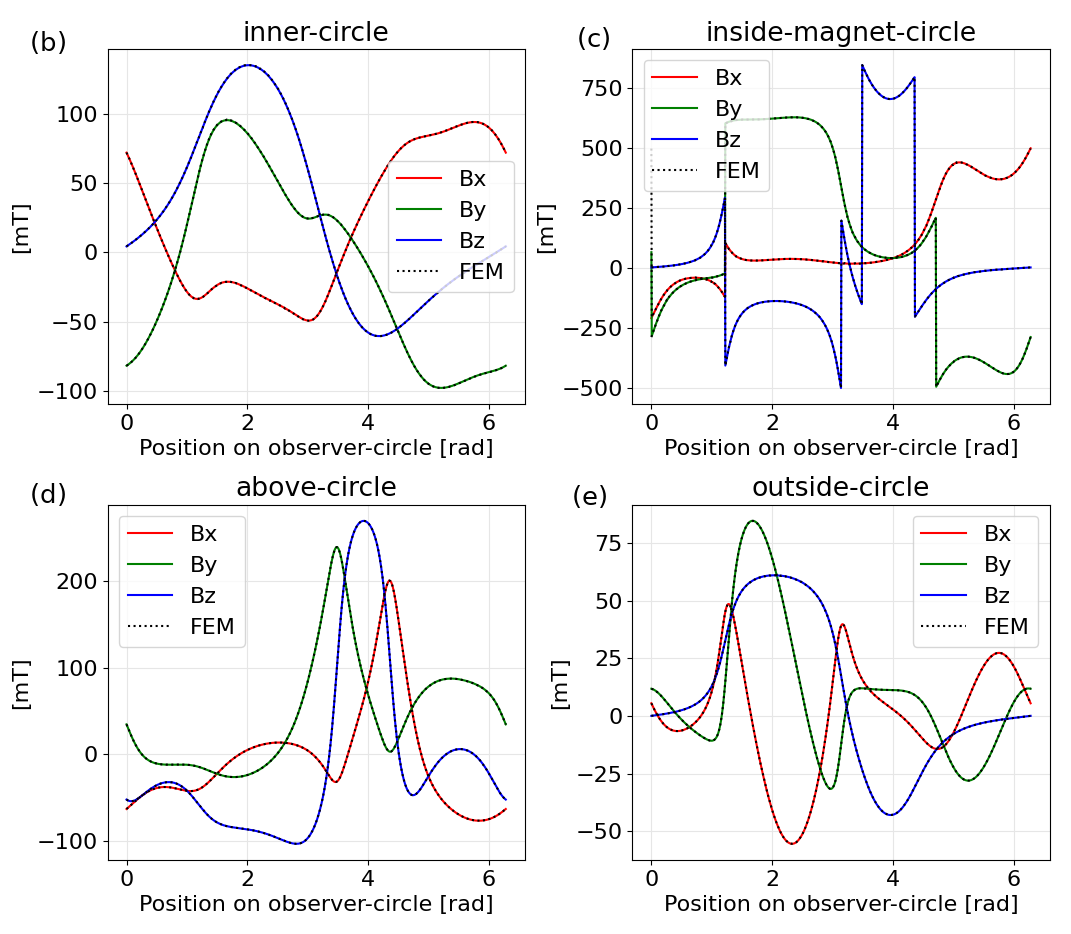}
    
    \end{subfigure}
    \caption{(a) Position of the three cylinder tiles in red, amber and green. The blue circles illustrate the paths, along which the magnetic field is evaluated. (b) Cartesian field components along the inner-circle. (c) Cartesian field components along the inside-magnet-circle. (d) Cartesian field components along the above-circle. (e) Cartesian field components along the outside-circle.}
    \label{fig::fields}
\end{figure}

Finally, we compare our Magpylib implementation with the state-of-the-art finite element computations from ANSYS Maxwell \citep{b72}. To this end, we place three different uniformly magnetized cylinder tiles with the dimensions given in Tab.~\ref{tab::tiles} in a circle, as illustrated in Fig.~\ref{fig::tiles}. The field is then computed on four rings, see Tab.~\ref{tab::circles}, above, outside, inside and on the inside of the tiles (blue in the figure) to cover many different special cases. The agreement between analytical implementation and the numerical evaluation can be observed in Figs.~\ref{fig::fields}b-e, and is only limited by the numerical precision of the finite element computation.
\begin{table}
\begin{center}
\begin{tabular}{ c||c|c|c|c|c|c|c|c|c } 
 & $r_1$ [mm] & $r_2$ [mm] & $\varphi_1$ [rad] & $\varphi_2$ [rad] & $z_1$ [mm] & $z_2$ [mm] & $\mu_0 M$ [T] & $\varphi_M$ [rad] & $\theta_M$ [rad] \\
 \hline
tile 1 & $1$ & $2$ & $3\pi/2$ & $2\pi$ & $-0.5$ & $0.5$ & $1$ &  $7\pi/4$ & $\pi/2$ \\
tile 2 & $1$ & $2.5$ & $10\pi/9$ & $25\pi/18$ & $-0.75$ & $0.75$ & $1$ & $0$ & $0$ \\
tile 3 & $0.75$ & $3$ & $7\pi/18$ & $\pi$ & $-0.25$ & $0.25$ & $1$ & $\pi/2$ & $3\pi/4$
\end{tabular}
\end{center}
\caption{Position and magnetization of the three cylinder tiles, as illustrated in Fig.~\ref{fig::tiles}.}
\label{tab::tiles}
\end{table}
\begin{table}
\begin{center}
\begin{tabular}{ c||c|c|c } 
 & center [mm] & radius [mm] & rotation axis [mm] \\
 \hline
inner-circle & $(0,0,0)$ & $0.5$ & $(0,0,1)$ \\
inside-magnet-circle & $(0,0,0)$ & $1.5$ & $(0,0,1)$ \\
above-circle & $(0,0,1)$ & $1.5$ & $(0,0,1)$ \\
outside-circle & $(0,0,0)$ & $3.5$ & $(0,0,1)$
\end{tabular}
\end{center}
\caption{Center, radius and rotation axis of the four circles along which the magnetic field is evaluated as illustrated in Fig.~\ref{tab::tiles}.}
\label{tab::circles}
\end{table}

\section{Halbach cylinder application}\label{sec::halbach}
\begin{figure}
    \centering
        \begin{subfigure}[b]{0.27\textwidth}
        \centering
        \includegraphics[width=1.0\textwidth]{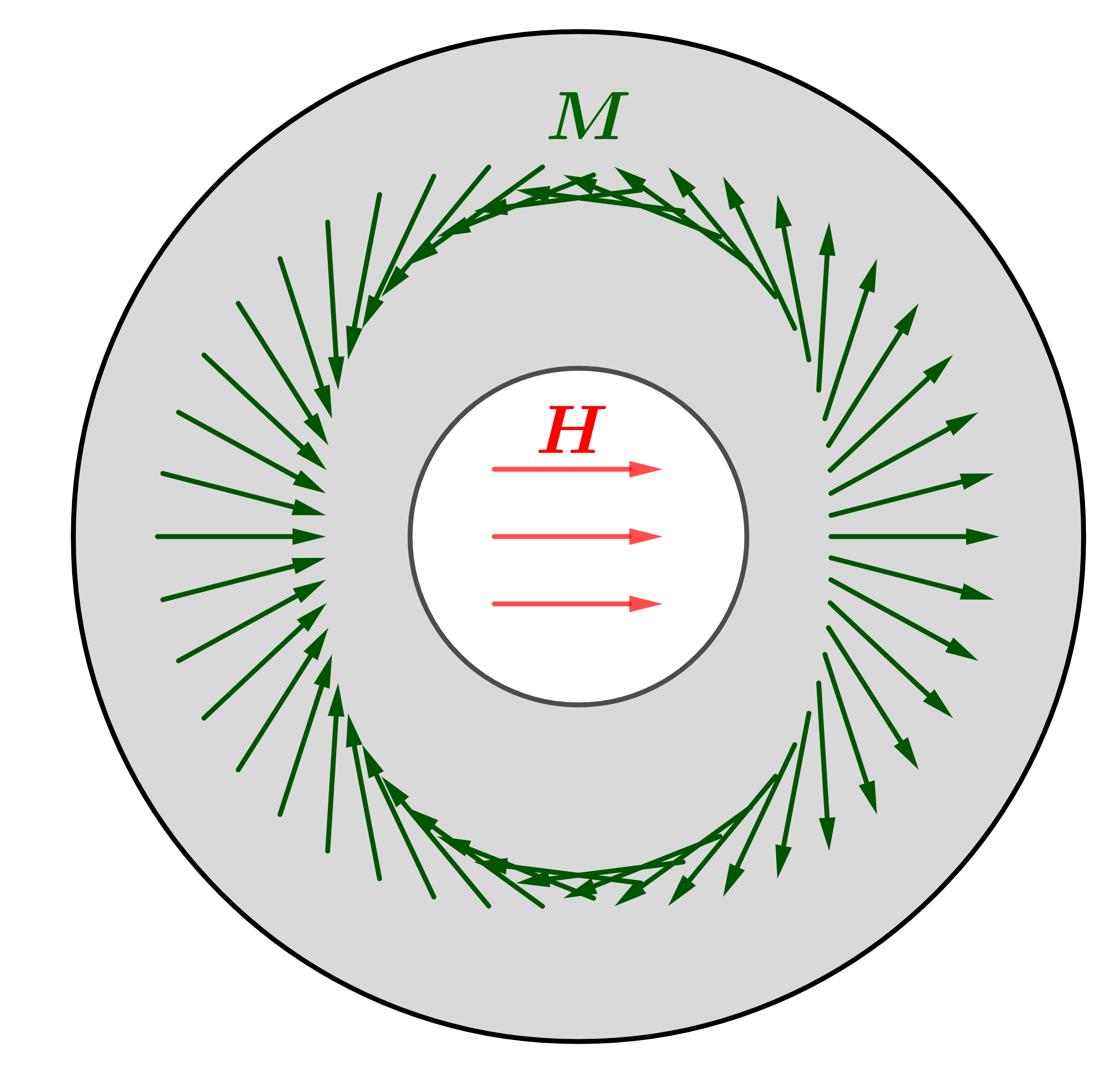}
        \caption{}. 
        \label{fig::halbach_cylinder}
    \end{subfigure}~\begin{subfigure}[b]{0.27\textwidth}
        \centering
        \includegraphics[width=1.0\textwidth]{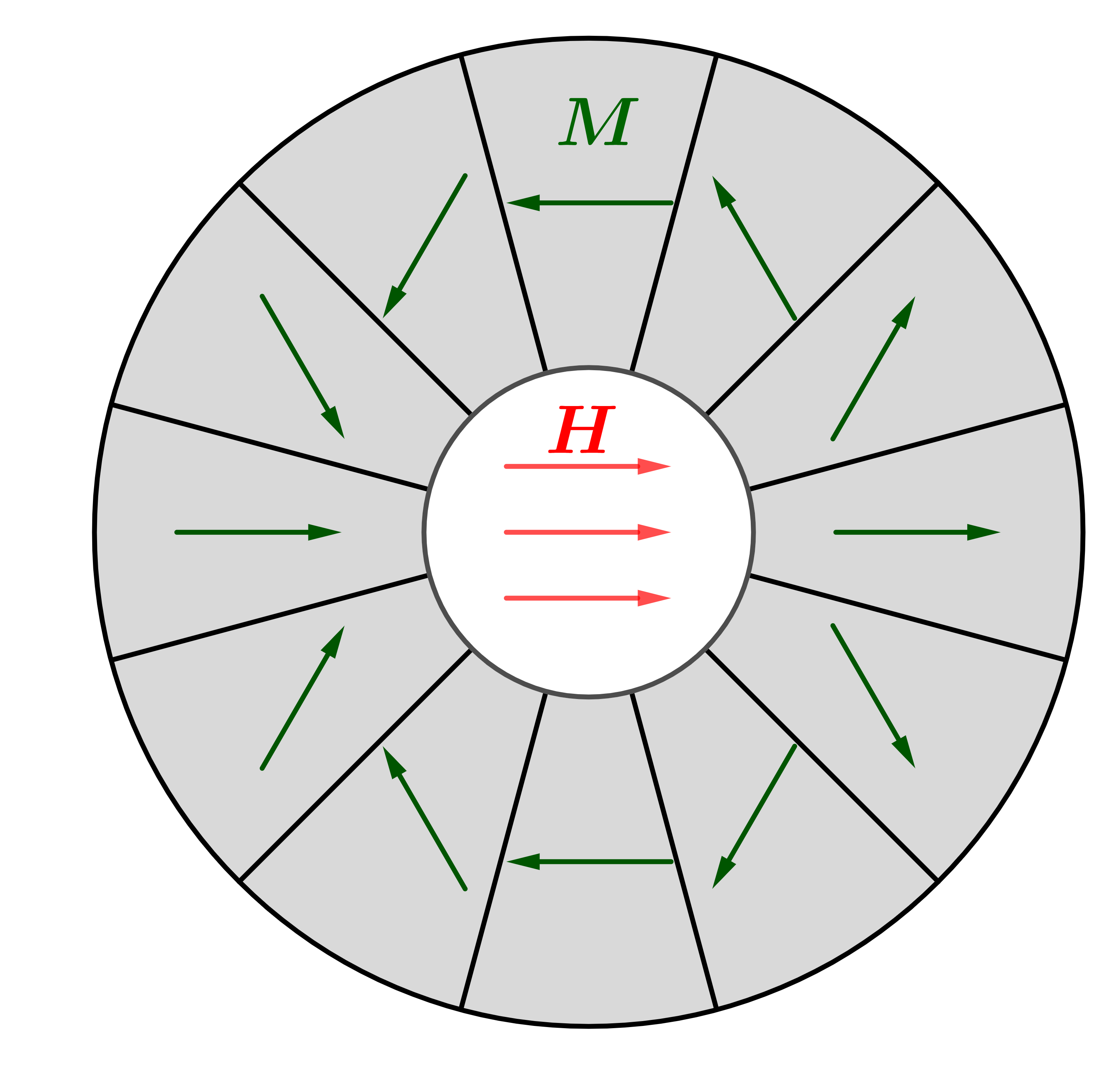}
        \caption{}. 
        \label{fig::halbach_cylinder_tiles}
    \end{subfigure}~\begin{subfigure}[b]{0.45\textwidth}
        \centering
        \includegraphics[width=1.0\textwidth]{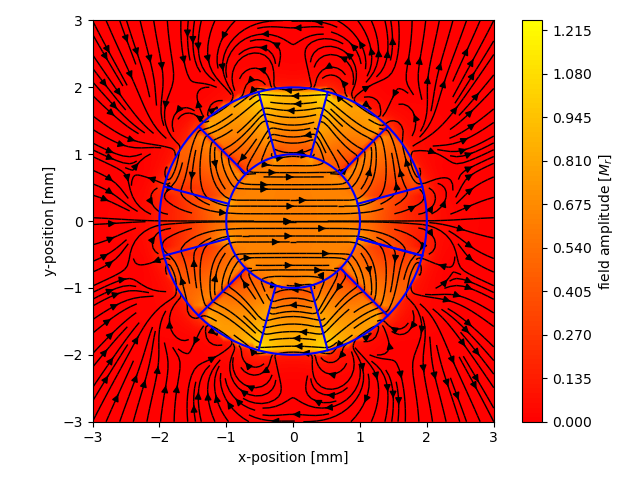}
        \caption{} 
        \label{fig::halbach}
    \end{subfigure}
    \caption{(a) Sketch of magnetization ${\bf M}$ in a Halbach cylinder with perfectly homogeneous field ${\bf H}$ inside. (b) Sketch of a discrete Halbach cylinder with $n=12$ cylinder tiles. (c) Magnetic flux density in the discrete Halbach cylinder with a height of 4~mm, computed with the analytical solution.}
\end{figure}

The computation of the field of discrete Halbach cylinders is a perfect use-case for our analytical computation method. Classical Halbach cylinders are magnetized hollow cylinders with a very specific azimuth angle dependent magnetization pattern,
\begin{align}
     {\bf M}_n (\varphi) = M_r (\cos(2m\varphi), \sin(2m\varphi),0), \quad m\in \mathbb{N}
     \label{equ::halbach}
\end{align}
that results for the case $m = 1$ in a perfectly homogeneous field on the inside of the cylinder and no field on the outside, see Fig.~\ref{fig::halbach_cylinder}. Halbach cylinders and their derivatives are magnetically the most efficient structures to generate homogeneous fields with large field amplitudes. They are commonly used for magnetic resonance imaging \citep{b73}, magnetic refrigeration \citep{b74}, for field annealing, energy harvesting \citep{b75} and many other modern applications.


For practical purposes, it is very difficult to achieve the magnetization pattern \eqref{equ::halbach}. However, a common approach replaces the continuous magnetization pattern with a discrete one, \citep{b73}. In Fig.~\ref{fig::halbach_cylinder_tiles} we show such a discretization, where a discrete Halbach cylinder is constructed from homogeneously magnetized cylinder tiles.


The analytical formulas presented in this work enable users to quickly compute and test the field of such discrete Halbach structures for different cylinder radii, cylinder heights, discretizations, magnetization patterns, target regions and all other variables that come into play when designing an experiment. A specific example for a discretization into $n=12$ segments is shown in Fig.~\ref{fig::halbach}.


\section{Conclusion \& outlook}\label{sec::outlook}
In this work, we have presented fully analytical expressions for the magnetic field of a homogeneously magnetized cylinder tile. The solution is implemented in Python and validated against numerical computations. The usefulness of this work is demonstrated by computing the field in a discrete Halbach cylinder.


While the provided formula set is complete from a mathematical point of view, floating point based implementations still suffer from numerical instabilities in the vicinity of special cases, indeterminate forms and at large distances. To address this problem, methods of reformulation and series expansion are planned to provide a higher level of stability.

By including the code in the open-source Python package Magpylib, the benefits of this work are also immediately accessible to the general public. Future work is dedicated to improve the Magpylib implementation work flow, stability and computation speed.


\section*{Acknowledgment}
This project has been supported by the COMET K1 centre ASSIC Austrian Smart Systems Integration 100 Research Center. The COMET --- Competence Centers for Excellent Technologies --- Program is supported by BMVIT, 101 BMDW and the federal provinces of Carinthia and Styria. Financial support by the Vienna Doctoral School in Physics (VDSP) is also acknowledged.

\appendix


\section{How to avoid integration singularities}\label{app::singularies}

\subsection{Motivation}
It is not surprising if singularities appear in the magnetic field ${\bf H}$ when crossing magnetically charged surfaces. However, it is also possible that single anti-derivatives contributing to the total field Eq.~\ref{eq::totalfield} have singularities off-surface. As the field itself must be continuous, these singularities cancel by summation with the singularities of other anti-derivatives.


For example, if we consider Eq.~\eqref{eq::rz} for the special case $z=z_k, r>0$, the term has a logarithmic singularity at $\varphi=\varphi'$ if $r>r_i$, which disappears after the following integration step. 
In contrast, for the case $r=r_i$ in Eq.~\eqref{eq::phir}, the denominator vanishes for $\overline{\varphi}' \in2\pi\mathbb Z$ and reveals a significant singularity that persists even after integration in the case $\overline{\varphi}_j\in 2\pi\mathbb Z$, making straight-forward evaluation of this term impossible.

It is crucial to understand how to eliminate the occurrence of these singularities, especially in the second integration step. Our study reveals that they are the result of poorly chosen integration constants by the computer algebra systems.
How this happens and how to avoid it will be shown in the following sections.

Appendix~\ref{sec::example2}). 
\subsection{Notation}
Let $A \subseteq \mathbb R^{n+2}$ be a domain for the function
\begin{align*}
f:& A \longrightarrow \mathbb R\\
&(x,y,p_1,\ldots,p_n) \longmapsto f(x,y,p_1,\ldots,p_n)
\end{align*}
in two variables $x,y$ and $n$ additional parameters $p_1, \ldots, p_n$. We are looking for an anti-derivative $F: A \longrightarrow \mathbb R$ satisfying $\partial_x \partial_y F = f = \partial_y \partial_x F$. It exists if $f$ is continuous and bounded, then $F$ does exist in the whole domain $A$ but is not unique. Suppose we compute $F$ in two steps (e.g. using a computer algebra system):
\begin{itemize}
    \item calculating an $x$-anti-derivative $F_x$ of $f$
    \item calculating an $y$-anti-derivative $F_{x,y}$ of $F_x$
\end{itemize}
The order of integration is mathematically irrelevant, but it should be noted that it can lead to different expressions when using computer algebra systems. This is demonstrated and explained in the following sections.

\subsection{Singularities in the integration constants}
Since in the calculation of $F_x = \int f\text{ d}x$ all variables except $x$ are fixed, the integration constant can be an arbitrary function of $(y, p_1, \ldots, p_n)$. This constant may have any number of singularities in $A$, so $F_x$ with integration constant may have a restricted domain $A' \subset A$. The same problem may arise in the second integration step, the integration of $F_x$ with integration constant by $y$. Again, another integration constant may occur depending on $(x, p_1, \ldots, p_n)$ with possible singularities in the result.

When integrating, computer algebra systems attempt to choose good integration constants. This possibility, together with the ambition to write results in ``simple'' and ``closed'' form, can result in adding singular integration constants. These unwanted singularities are hidden in the resulting deceptively simple looking expressions. We illustrate this with examples in the following subsections.

\subsection{Example 1}
The function
\begin{align*}
    f(x,p) := \frac{1}{\sqrt{x^2+p}}
\end{align*}
on the domain $\{(x,p)\in\mathbb R^2 \big\vert p\geq 0, (x,p) \neq (0,0)\}$ is continuous and bounded if one removes a small region around $(0,0)$, so the anti-derivative $F$ must exist for all $p$ in the domain. The integration of $f$ by $x$ can lead to one of the following three anti-derivatives in computer algebra systems:
\begin{align*}
    F_{x,1}(x,p) &= \log\left(\sqrt{x^2+p} + x\right) \\
    F_{x,2}(x,p) &= -\log\left(\sqrt{x^2+p} - x\right) \\
    F_{x,3}(x,p) &= \artanh{\left( \frac{x}{\sqrt{x^2+p}} \right)} \\
\end{align*}

Mathematica version 10.1 \citep{mathematica} for instance produces $F_{x,1}$ with its standard integrator. The additionally loaded Rubi integrator \citep{rubi} results in $F_{x,3}$.

It is easy to see that all three functions are well-defined in the case $p>0$ for all $x$. However, for $p=0$ we see that $F_{x,1}$ is defined only for $x>0$, $F_{x,2}$ is defined only for $x<0$ and $F_{x,3}$ is not defined for any $x\in \mathbb R$ since the domain of $\artanh$ is $(-1, 1)$. Further computations show that the three possible outputs of a computer algebra system, although looking different at first glance, differ only by a $p$-dependent integration constant, which has a singularity at $p=0$, i.e. $F_{x,1}(x,p) - F_{x,2}(x,p) = \log(p)$ and $F_{x,3}(x,p) - F_{x,2}(x,p) = \log(p) / 2$. As described above, we observe an implicit integration constant that reduces the domain for which our anti-derivative can be used. The automatic simplification of the term to known mathematical function as $\log, \artanh$ covers this problem.

\subsection{Example 2}\label{sec::example2}
We now study an example function similar to $H_{\varphi,r_i}$ in Eq.~\eqref{eq::Hphi} and its anti-derivative in Eq.~\eqref{eq::phir}:
\begin{align*}
    f(\varphi,z) := \frac{\sin\varphi}{\sqrt{1-\cos\varphi + z^2}^3}
\end{align*}
on the domain 
\begin{align}\label{eq::domain}
\{(\varphi,z)\in [0,2\pi) \times \mathbb R \big\vert (\varphi,z) \neq (0,0) \}.
\end{align}

\begin{figure}
    \centering
    \includegraphics[width=0.6\textwidth]{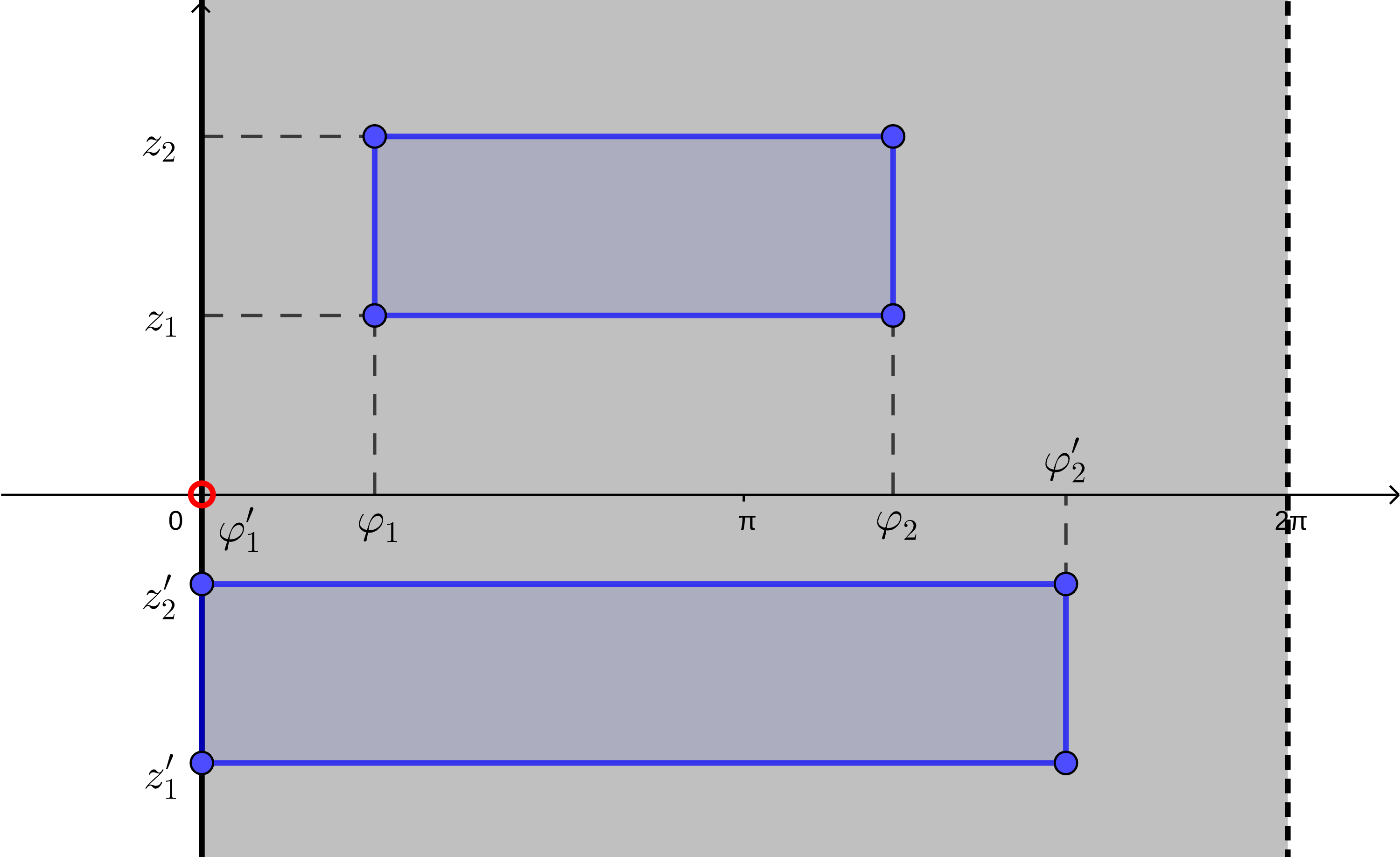}
    \caption{The upper rectangle represents the integration area $[\varphi_1,\varphi_2] \times [z_1,z_2]$. In this case, $z_1$ could be moved below the horizontal axis without leaving the grayed region, i.e. the sign of $z_1$ could also be changed. However, the lower rectangle would intersect the point $(0,0)$ if the upper boundary $z_2$ is moved to the positive part of the vertical axis, since $\varphi'_1 = 0$. }
    \label{fig::domain}
\end{figure}

For $z\neq 0$, the function term is obviously well-defined for all $\varphi$, since $(1-\cos\varphi) \geq 0$. For $z=0$, however, L'Hospital's rule reveals significant singularities,
\begin{align*}
    \lim_{\varphi\rightarrow 0} \frac{\sin\varphi}{\sqrt{1-\cos\varphi}^3} = \lim_{\varphi\rightarrow 0} \frac{\cos\varphi}{3/2\sqrt{1-\cos\varphi}\sin\varphi} = \frac{1}{0}.
\end{align*}
An anti-derivative of $f$ by $z$ is given by
\begin{align*}
    F_{z,1}(\varphi,z) = \frac{z \sin\varphi}{(1-\cos\varphi)\sqrt{1-\cos\varphi + z^2}},
\end{align*}
however, the additional factor $(1-\cos\varphi)$ in the denominator also generates a pole for $z\neq 0$ at $\varphi=0$, which again can be seen with L'Hospital's rule,
\begin{align*}
    \lim_{\varphi\rightarrow 0} \frac{\sin\varphi}{1-\cos\varphi} = \lim_{\varphi\rightarrow 0} \frac{\cos\varphi}{\sin\varphi} = \frac{1}{0}.
\end{align*}
Thus $F_{z.1}$ can be used only for the case $\varphi \neq 0$. The same applies to the following anti-derivation by $\varphi$
\begin{align*}
    F_{z,\varphi, 1}(\varphi,z) = -2\artanh\left( \frac{z}{\sqrt{1 - \cos\varphi + z^2}} \right),
\end{align*}
which is again undefined in the case $\varphi = 0$ for all $z\in\mathbb R$. In order to obtain a proper anti-derivative also for the case $\varphi = 0$, we first note that our goal is to calculate definite integrals of the form

\begin{align*}
    \int_{\varphi_1}^{\varphi_2} \int_{z_1}^{z_2} f(\varphi,z)\text{ d}z\text{ d}\varphi.
\end{align*}
It is justified to restrict our domain of interest to a rectangular subset of Eq.~\eqref{eq::domain}. That is, if $\varphi = 0$ lies in the interval $[\varphi_1, \varphi_2]$, then we can assume that $z=0$ is not in $[z_1, z_2]$ since $(0,0)$ is excluded from the domain, and moreover that $z_1, z_2$ are either both positive or both negative, see Fig.~\ref{fig::domain}.

This argument allows us to add to $F_{z,1}$ a sign-dependent integration constant that also depends on $\varphi$ and compensates the pole for $\varphi=0$. We define
\begin{align*}
    F_{z,2}(\varphi,z) := \frac{\sin\varphi}{1-\cos\varphi} \left(\frac{z}{\sqrt{1-\cos\varphi + z^2}} - \sign z \right),
\end{align*}
where the term in parentheses vanishes for $\varphi=0$ and thus compensates the pole of the prefactor. Integration of this function by $\varphi$ gives
\begin{align*}
    F_{z,\varphi,2}(\varphi,z) = -2\sign z \log\left( \sqrt{1-\cos\varphi + z^2} + |z| \right).
\end{align*}
which has no singularity in the case $\varphi = 0, z\neq 0$ any more, and can be used to compute the definite integral when $\varphi_1 = 0$ or $\varphi_2 = 0$.

\section{Special functions}\label{app::elliptic}
\subsection{Elliptic integrals}
In accordance with \citep[Chapter 8]{b2} we define the elliptic integrals in the angular form: 
\begin{itemize}
    \item First kind:
    \begin{align}
        F(\varphi, m)=\int_{0}^{\varphi} \frac{\text{d} \varphi'}{\sqrt{1-m \sin ^{2} \varphi'}}
    \end{align}
    \item Second kind:
    \begin{align}
        E(\varphi, m)=\int_{0}^{\varphi} \sqrt{1-m \sin ^{2} \varphi'}\ \text{d} \varphi'
    \end{align}
    \item Third kind:
    \begin{align}
        \Pi(\varphi, n, m)=\int_{0}^{\varphi} \frac{\text{d} \varphi'}{\left(1-n \sin ^{2} \varphi'\right) \sqrt{1-m \sin ^{2} \varphi'}}
    \end{align}
\end{itemize}
These integrals exist and yield real numbers for $\varphi\in\mathbb R$ and $m,n\in (-\infty,1)$. For $\varphi=\pi/2$ they are called \textit{complete}, otherwise \textit{incomplete} elliptic integrals. There are several very efficient numerical algorithms to compute these integrals with effective computation times between 10~ns and 1~$\mu$s for single-core evaluation on state-of-the-art i5 or i7 mobile CPUs \citep{b35,b5,b6,b7,b8}.

\subsection{Angle scaling function}
\begin{figure}
    \centering
    \includegraphics[width=0.6\textwidth]{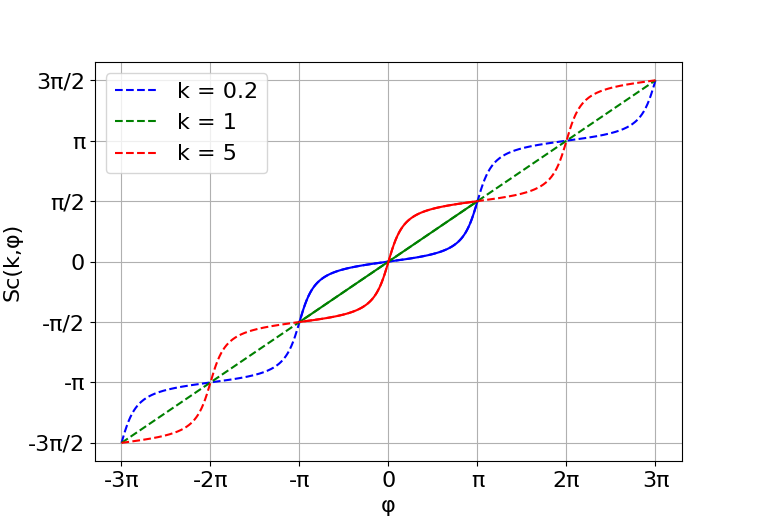}
    \caption{Graph of the angle scaling function $\Sc$ as the periodic continuous continuation of the anti-derivative from $(-\pi,\pi)$ to $\mathbb R$ for three different values of $k$. }
    \label{fig::sc}
\end{figure}

For some results in Appendix~\ref{app::tables}, we make use of the following angle scaling function
\begin{align*}
\Sc(\varphi, k) := \pi n + \arctan(k\tan(\varphi/2 - \pi n)),
\end{align*}
where $n:=[ \varphi/2\pi ]$, i.e. the integer value closest to $\varphi/2\pi$, which is a periodic continuation of the anti-derivative $\arctan(k\tan(\varphi/2))$ with $k>0$ (see Fig.~\ref{fig::sc}).

\begin{landscape}

\section{Tables}\label{app::tables}
In Tabs.~\ref{tab::case112}-\ref{tab::case235} all necessary functions with corresponding coefficients to perform the field calculation according to Sec.~\ref{sec::2ndint} are given. How the tables are to be read is shown in Tab.~\ref{tab::examplecase}. All terms with $\pm$-sign must be considered twice, with plus and minus.

\begin{table}
\begin{center}
\begin{tabular}{ c||c|c } 
& $1$ & $\log{r_i}$ \\
 \hline\hline
 $H_{\varphi,z_k}^{\varphi_j,r_i}$ & $0$ & $\cos{\theta_M}$ \\ 
 \hline
 $H_{z,r_i}^{z_k,\varphi_j}$ & $-\sin{\theta_M}\sin{\overline{\varphi}_M}$ & $0$ \\
 \hline
 $H_{z,\varphi_j}^{z_k,r_i}$ & $0$ & $\sin{\theta_M}\sin{\overline{\varphi}_M}$  
\end{tabular}
\end{center}
\caption{Required functions and corresponding coefficients for $\mathcal{I} = 112$.}
\label{tab::case112}
\end{table}

\begin{table}
\begin{center}
\begin{tabular}{ c||c } 
&  $\log{r}$ \\
 \hline\hline
 $H_{\varphi,z_k}^{\varphi_j,r_i}$ & $-\cos{\theta_M}$ \\ 
 \hline
 $H_{z,\varphi_j}^{z_k,r_i}$ & $-\sin{\theta_M}\sin{\overline{\varphi}_M}$  
\end{tabular}
\end{center}
\caption{Required functions and corresponding coefficients for $\mathcal{I} = 113$.}
\label{tab::case113}
\end{table}

\begin{table}
\begin{center}
\begin{tabular}{ c||c|c|c|c } 
& $1$ & $\sign{\overline{r}_i} \log\left|\overline{r}_i\right|$ & $E\left(\frac{\overline{\varphi}_j}{2}, -\frac{4rr_i}{\overline{r}_i^2}\right)$ &  $F\left(\frac{\overline{\varphi}_j}{2}, -\frac{4rr_i}{\overline{r}_i^2}\right)$ \\
 \hline\hline
 $H_{r,z_k}^{r_i,\varphi_j}$ & $0$ & $0$ & $\frac{\cos \theta_M \left|\overline{r}_i\right|}{r}$ & $-\frac{\cos \theta_M \left(r^2 + r_i^2\right)}{r \left|\overline{r}_i\right|}$ \\ 
 \hline
 $H_{\varphi,z_k}^{\varphi_j,r_i}$ & $-\frac{\cos \theta_M \sign{\overline{r}_i r_i}}{r}$ & $-\cos \theta_M $ & $0$ & $0$ \\
 \hline
 $H_{z,r_i}^{z_k,\varphi_j}$ & $\frac{\sin \theta_M \sin \overline{\varphi}_M\left|\overline{r}_i\right|}{r}$ & $0$ & $\frac{\sin \theta_M \cos \overline{\varphi}_M \left|\overline{r}_i\right|}{r}$ & $-\frac{\sin \theta_M \cos \overline{\varphi}_M \left(r^2 + r_i^2\right)}{r \left|\overline{r}_i\right|}$ \\
 \hline
 $H_{z,\varphi_j}^{z_k,r_i}$ & $0$ & $-\sin \theta_M \sin \overline{\varphi}_M$ & $0$ & $0$
\end{tabular}
\end{center}
\caption{Required functions and corresponding coefficients for $\mathcal{I} = 115$.}
\label{tab::case115}
\end{table}

\begin{table}
\begin{center}
\begin{tabular}{ c||c|c } 
& $1$ & $\log{r_i}$ \\
 \hline\hline
 $H_{\varphi,z_k}^{\varphi_j,r_i}$ & $0$ & $-\cos{\theta_M}$ \\ 
 \hline
 $H_{z,r_i}^{z_k,\varphi_j}$ & $\sin{\theta_M}\sin{\overline{\varphi}_M}$ & $0$ \\ 
 \hline
 $H_{z,\varphi_j}^{z_k,r_i}$ & $0$ & $-\sin{\theta_M}\sin{\overline{\varphi}_M}$  
\end{tabular}
\end{center}
\caption{Required functions and corresponding coefficients for $\mathcal{I} = 122$.}
\label{tab::case122}
\end{table}

\begin{table}
\begin{center}
\begin{tabular}{ c||c } 
&  $\log{r}$ \\
 \hline\hline
 $H_{\varphi,z_k}^{\varphi_j,r_i}$ & $-\cos{\theta_M}$ \\ 
 \hline
 $H_{z,\varphi_j}^{z_k,r_i}$ & $-\sin{\theta_M}\sin{\overline{\varphi}_M}$  
\end{tabular}
\end{center}
\caption{Required functions and corresponding coefficients for $\mathcal{I} = 123$.}
\label{tab::case123}
\end{table}

\begin{table}
\begin{center}
\begin{tabular}{ c||c|c } 
& $1$ & $\log{2r}$ \\
 \hline\hline
 $H_{\varphi,z_k}^{\varphi_j,r_i}$ & $\cos{\theta_M}$ & $-\cos{\theta_M}$ \\ 
 \hline
 $H_{z,r_i}^{z_k,\varphi_j}$ & $2\sin{\theta_M}\sin\overline{\varphi}_M$ & $0$ \\ 
 \hline
 $H_{z,\varphi_j}^{z_k,r_i}$ & $0$ & $-\sin{\theta_M}\sin{\overline{\varphi}_M}$  
\end{tabular}
\end{center}
\caption{Required functions and corresponding coefficients for $\mathcal{I} = 124$.}
\label{tab::case124}
\end{table}

\begin{table}
\begin{center}
\begin{tabular}{ c||c|c|c|c } 
& $1$ & $\log\left(r+r_i\right)$ & $E\left(\frac{\overline{\varphi}_j}{2}, -\frac{4rr_i}{\overline{r}_i^2}\right)$ &  $F\left(\frac{\overline{\varphi}_j}{2}, -\frac{4rr_i}{\overline{r}_i^2}\right)$ \\
 \hline\hline
 $H_{r,z_k}^{r_i,\varphi_j}$ & $0$ & $0$ & $\frac{\cos \theta_M \left|\overline{r}_i\right|}{r}$ & $-\frac{\cos \theta_M \left(r^2 + r_i^2\right)}{r \left|\overline{r}_i\right|}$ \\ 
 \hline
 $H_{\varphi,z_k}^{\varphi_j,r_i}$ & $\frac{\cos \theta_M r_i}{r}$ & $-\cos \theta_M $ & $0$ & $0$ \\
 \hline
 $H_{z,r_i}^{z_k,\varphi_j}$ & $\frac{\sin\theta_M \sin\overline{\varphi}_M\left( r + r_i \right)}{r}$ & $0$ & $\frac{\sin \theta_M \cos \overline{\varphi}_M \left|\overline{r}_i\right|}{r}$ & $-\frac{\sin \theta_M \cos \overline{\varphi}_M \left(r^2 + r_i^2\right)}{r \left|\overline{r}_i\right|}$ \\
 \hline
 $H_{z,\varphi_j}^{z_k,r_i}$ & $0$ & $-\sin \theta_M \sin \overline{\varphi}_M$ & $0$ & $0$
\end{tabular}
\end{center}
\caption{Required functions and corresponding coefficients for $\mathcal{I} = 125$.}
\label{tab::case125}
\end{table}

\begin{table}
\begin{center}
\begin{tabular}{ c||c|c } 
& $1$ & $\log{r_i}$ \\
 \hline\hline
 $H_{r,z_k}^{r_i,\varphi_j}$ & $0$ & $\cos{\theta_M} \sin\overline{\varphi}_j$ \\
 \hline
 $H_{\varphi,z_k}^{\varphi_j,r_i}$ & $0$ & $\cos{\theta_M}\cos\overline{\varphi}_j$ \\ 
 \hline
 $H_{z,r_i}^{z_k,\varphi_j}$ & $-\sin{\theta_M}\sin{\overline{\varphi}_{Mj}}$ & $0$ \\ 
 \hline
 $H_{z,\varphi_j}^{z_k,r_i}$ & $0$ & $\sin{\theta_M}\sin{\overline{\varphi}_{Mj}}$  
\end{tabular}
\end{center}
\caption{Required functions and corresponding coefficients for $\mathcal{I} = 132$.}
\label{tab::case132}
\end{table}

\begin{table}
\begin{center}
\begin{tabular}{ c||c|c|c } 
& $1$ & $\log{\left(r\left(1-\cos\overline{\varphi}_j\right)\right)}$ & $\artanh\left(\cos\overline{\varphi}_j\right)$ \\
 \hline\hline
  $H_{r,z_k}^{r_i,\varphi_j}$ & 
  $-\cos{\theta_M} \sin\overline{\varphi}_j$ & 
  $\cos{\theta_M} \sin\overline{\varphi}_j$ & 
  $0$
  \\
  \hline
 $H_{\varphi,z_k}^{\varphi_j,r_i}$ & 
 $\cos{\theta_M}$ & 
 $0$ & 
 $-\cos{\theta_M} \cos\overline{\varphi}_j$ 
 \\ 
 \hline
 $H_{z,\varphi_j}^{z_k,r_i}$ & 
 $0$ & 
 $0$ & 
 $-\sin{\theta_M}\sin{\overline{\varphi}_{Mj}}$  
\end{tabular}
\end{center}
\caption{Required functions and corresponding coefficients for $\mathcal{I} = 133$.}
\label{tab::case133}
\end{table}

\begin{table}
\begin{center}
\begin{tabular}{ c||c|c|c|c|c } 
& $1$ & $\sqrt{2}\sqrt{1 - \cos\overline{\varphi}_j}$ & $\artanh{\left( \sqrt{\frac{1 - \cos\overline{\varphi}_j}{2}} \right)}$ & $\artanh{\left( \frac{\sin\overline{\varphi}_j}{\sqrt{2}\sqrt{1 - \cos\overline{\varphi}_j}} \right)} - \frac{\sqrt{2}\sin\overline{\varphi}_j}{\sqrt{1 - \cos\overline{\varphi}_j}}$ & $\log\left( r\left(1 - \cos\overline{\varphi}_j + \sqrt{2}\sqrt{1 - \cos\overline{\varphi}_j}\right) \right)$ \\
 \hline\hline
 $H_{r,z_k}^{r_i,\varphi_j}$ & $-\cos{\theta_M}\sin{\overline{\varphi}_{j}}$ & $0$  & $0$ & $\cos{\theta_M}$ & $\cos{\theta_M}\sin{\overline{\varphi}_{j}}$ \\
 \hline
 $H_{\varphi,z_k}^{\varphi_j,r_i}$ & $0$ & $\cos{\theta_M}$ & $\cos{\theta_M} \cos\overline{\varphi}_j$ & $0$ & $0$ \\ 
 \hline
 $H_{z,r_i}^{z_k,\varphi_j}$ & $0$ & $\sin\theta_M\sin{\overline{\varphi}_{M}}$ & $0$ & $\sin\theta_M\cos{\overline{\varphi}_{M}}$ & $0$ \\ 
 \hline
 $H_{z,\varphi_j}^{z_k,r_i}$ & $0$ & $0$ & $\sin{\theta_M} \sin\overline{\varphi}_{Mj}$ & $0$ & $0$
\end{tabular}
\end{center}
\caption{Required functions and corresponding coefficients for $\mathcal{I} = 134$.}
\label{tab::case134}
\end{table}

\begin{table}
\begin{center}
\begin{tabular}{ c||c|c|c|c|c|c } 
& $1$ & $\frac{\sqrt{r^2 + r_i^2 - 2rr_i\cos{\overline{\varphi}_{j}}}}{r}$ & $\log\left(r_i - r\cos{\overline{\varphi}_{j}} + \sqrt{r^2 + r_i^2 - 2rr_i\cos{\overline{\varphi}_{j}}}\right)$ & $\artanh{\left( \frac{r\cos{\overline{\varphi}_{j}} - r_i}{\sqrt{r^2 + r_i^2 - 2rr_i\cos{\overline{\varphi}_{j}}}} \right)}$ & $E\left(\frac{\overline{\varphi}_j}{2}, -\frac{4rr_i}{\overline{r}_i^2}\right)$ &  $F\left(\frac{\overline{\varphi}_j}{2}, -\frac{4rr_i}{\overline{r}_i^2}\right)$ \\
 \hline\hline
 $H_{r,z_k}^{r_i,\varphi_j}$ & $-\cos \theta_M\sin{\overline{\varphi}_{j}}$ & $0$ & $\cos \theta_M\sin{\overline{\varphi}_{j}}$ & $0$ & $\frac{\cos \theta_M \left|\overline{r}_i\right|}{r}$ & $-\frac{\cos \theta_M \left(r^2 + r_i^2\right)}{r \left|\overline{r}_i\right|}$ \\ 
 \hline
 $H_{\varphi,z_k}^{\varphi_j,r_i}$ & $0$ & $\cos \theta_M$ & $0$ & $-\cos \theta_M \cos{\overline{\varphi}_{j}} $ & $0$ & $0$ \\
 \hline
 $H_{z,r_i}^{z_k,\varphi_j}$ & $0$ & $\sin \theta_M \sin \overline{\varphi}_M$ & $0$ & $0$ & $\frac{\sin \theta_M \cos \overline{\varphi}_M \left|\overline{r}_i\right|}{r}$ & $-\frac{\sin \theta_M \cos \overline{\varphi}_M \left(r^2 + r_i^2\right)}{r \left|\overline{r}_i\right|}$ \\
 \hline
 $H_{z,\varphi_j}^{z_k,r_i}$ & $0$ & $0$ & $0$ & $-\sin \theta_M \sin \overline{\varphi}_{Mj}$ & $0$ & $0$
\end{tabular}
\end{center}
\caption{Required functions and corresponding coefficients for $\mathcal{I} = 135$.}
\label{tab::case135}
\end{table}

\begin{table}
\begin{center}
\begin{tabular}{ c||c|c } 
& $\sign{\overline{z}_k}$ & 
$\sign{\overline{z}_k} \log \left|\overline{z}_k\right|$ 
\\
 \hline\hline
 $H_{r,\varphi_j}^{r_i,z_k}$ & 
 $0$ & 
 $-\sin{\theta_M} \sin{\overline{\varphi}_M}$ 
 \\ 
 \hline
 $H_{z,z_k}^{r_i,\varphi_j}$ & 
 $-\cos{\theta_M} \varphi_j$ & 
 $0$
\end{tabular}
\end{center}
\caption{Required functions and corresponding coefficients for $\mathcal{I} = 211$.}
\label{tab::case211}
\end{table}

\begin{table}
\begin{center}
\begin{tabular}{ c||c|c|c|c } 
& $\frac{r_i}{\sqrt{r_i^2 + \overline{z}_k^2}}$ & 
$\artanh{\left( \frac{r_i}{\sqrt{r_i^2 + \overline{z}_k^2}} \right)}$ &
$\frac{\overline{z}_k}{\sqrt{r_i^2 + \overline{z}_k^2}}$ & 
$\artanh{\left( \frac{\overline{z}_k}{\sqrt{r_i^2 + \overline{z}_k^2}} \right)}$ 
\\
 \hline\hline
 $H_{r,r_i}^{z_k,\varphi_j}$ & 
 $0$ & 
 $0$ & 
 $\sin{\theta_M} \left( \frac{\cos{\overline{\varphi}_M}\varphi_j}{2} - \frac{\sin{\overline{\varphi}_M}}{4} \right)$ & 
 $0$ 
 \\
 \hline
 $H_{r,\varphi_j}^{r_i,z_k}$ & 
 $0$ & 
 $0$ & 
 $0$ & 
 $-\sin{\theta_M}\sin{\overline{\varphi}_M}$  
 \\
 \hline
 $H_{\varphi,r_i}^{z_k,\varphi_j}$ & 
 $0$ & 
 $0$ & 
 $\sin{\theta_M} \left( \frac{\sin{\overline{\varphi}_M}\varphi_j}{2} - \frac{\cos{\overline{\varphi}_M}}{4} \right)$ & 
 $0$
 \\
 \hline
 $H_{\varphi,z_k}^{\varphi_j,r_i}$ & 
 $-\cos{\theta_M}$ & 
 $\cos{\theta_M}$ & 
 $0$ & 
 $0$ 
 \\ 
 \hline
 $H_{z,r_i}^{z_k,\varphi_j}$ & 
 $-\sin{\theta_M}\sin{\overline{\varphi}_M}$ & 
 $0$ & 
 $0$ & 
 $0$ 
 \\ 
 \hline
 $H_{z,\varphi_j}^{z_k,r_i}$ & 
 $0$ & 
 $\sin{\theta_M}\sin{\overline{\varphi}_M}$ & 
 $0$ & 
 $0$ 
 \\
 \hline
  $H_{z,z_k}^{r_i,\varphi_j}$ & 
  $0$ & 
  $0$ & 
  $-\cos{\theta_M} \varphi_j$ & 
  $0$
\end{tabular}
\end{center}
\caption{Required functions and corresponding coefficients for $\mathcal{I} = 212$.}
\label{tab::case212}
\end{table}

\begin{table}
\begin{center}
\begin{tabular}{ c||c|c|c|c|c } 
& $\sign{\overline{z}_k}$ &
$\frac{\sqrt{r^2 + \overline{z}_k^2}}{r}$ & 
$\artanh{\left( \frac{r}{\sqrt{r^2 + \overline{z}_k^2}} \right)}$ & 
$\artanh{\left( \frac{\overline{z}_k}{\sqrt{r^2 + \overline{z}_k^2}} \right)}$ 
\\
 \hline\hline
 $H_{r,\varphi_j}^{r_i,z_k}$ & 
 $0$ &
 $0$ & 
 $0$ & 
 $-\sin{\theta_M}\sin{\overline{\varphi}_M} $ 
 \\
 \hline
 $H_{\varphi,z_k}^{\varphi_j,r_i}$ & 
 $0$ &
 $\cos{\theta_M}$ & 
 $-\cos{\theta_M}$ & 
 $0$ 
 \\ 
 \hline
 $H_{z,\varphi_j}^{z_k,r_i}$ & 
  $0$ & 
 $0$ & 
 $-\sin{\theta_M}\sin{\overline{\varphi}_M}$ & 
 $0$
 \\
 \hline
 $H_{z,z_k}^{r_i,\varphi_j}$ & 
 $\cos{\theta_M}\overline{\varphi}_j$ &
 $0$ & 
 $0$ & 
 $0$ 
\end{tabular}
\end{center}
\caption{Required functions and corresponding coefficients for $\mathcal{I} = 213$.}
\label{tab::case213}
\end{table}

\begin{table}
\begin{center}
\begin{tabular}{ c||c|c|c|c|c|c|c|c } 
& $\sign{\overline{z}_k}$ & 
$\frac{\left|\overline{z}_k\right|}{r}$ & 
$\frac{\sign{\overline{z}_k}\overline{z}_k^2}{2r^2}$ & 
$\sign{\overline{z}_k} \log \left|\overline{z}_k\right|$ & 
$\sign{\overline{z}_k}\log\left( \frac{\left|\overline{z}_k\right|}{\sqrt{2}r}\right)$ &
$E\left(\frac{\overline{\varphi}_j}{2}, -\frac{4r^2}{\overline{z}_k^2}\right)$ & 
$F\left(\frac{\overline{\varphi}_j}{2}, -\frac{4r^2}{\overline{z}_k^2}\right)$ & 
$\cdots$
\\
 \hline\hline
  $H_{r,r_i}^{z_k,\varphi_j}$ & 
  $0$ & 
  $0$ & 
  $-\sin{\theta_M}\sin{\overline{\varphi}_M}$ & 
  $0$ & 
  $0$ & 
  $-\frac{\sin{\theta_M}\cos{\overline{\varphi}_M} \sign{\overline{z}_k}\overline{z}_k^2}{2r^2}$ & 
  $\frac{\sin{\theta_M}\cos{\overline{\varphi}_M} \sign{\overline{z}_k} \left(2r^2 + \overline{z}_k^2\right)}{2r^2}$ & 
  \\ 
  \hline
 $H_{r,\varphi_j}^{r_i,z_k}$ & 
 $0$ & 
 $0$ & 
 $0$ & 
 $-\sin{\theta_M}\sin{\overline{\varphi}_M} $ & 
 $0$ & 
 $0$ & 
 $0$ & 
 \\
 \hline
 $H_{r,z_k}^{r_i,\varphi_j}$ & 
 $0$ & 
 $0$ & 
 $0$ & 
 $0$ & 
 $0$ & 
 $\frac{\cos{\theta_M}\left|\overline{z}_k\right|}{r}$ &
 $-\frac{\cos{\theta_M}\left(2r^2 + \overline{z}_k^2\right)}{r \left|\overline{z}_k\right|}$ &
$\cdots$
 \\
 \hline
  $H_{\varphi,r_i}^{z_k,\varphi_j}$ & 
  $\frac{\sin{\theta_M} \left(\sin{\overline{\varphi}_M}\varphi_j - \cos{\overline{\varphi}_M}\right)}{2}$ & 
  $0$ & 
  $-\sin{\theta_M} \cos{\overline{\varphi}_M}$ & 
  $0$ & 
  $-\sin{\theta_M} \cos{\overline{\varphi}_M}$ & 
  $\frac{\sin{\theta_M}\sin{\overline{\varphi}_M} \sign{\overline{z}_k}\overline{z}_k^2}{2r^2}$ & 
  $-\frac{\sin{\theta_M}\sin{\overline{\varphi}_M} \sign{\overline{z}_k} \left(4r^2 + \overline{z}_k^2\right)}{2r^2}$ & 
  \\
  \hline
 $H_{\varphi,z_k}^{\varphi_j,r_i}$ & 
 $0$ & 
 $\cos{\theta_M}$ & 
 $0$ & 
 $0$ & 
 $0$ & 
 $0$ & 
 $0$ & 
 \\
\end{tabular}

\begin{tabular}{ c||c|c|c } 
& $\cdots$ &
$\Pi\left( \frac{\overline{\varphi}_j}{2}, \frac{2r}{r\pm\sqrt{r^2 + \overline{z}_k^2}}, -\frac{4r^2}{\overline{z}_k^2} \right)$ & 
$\Pi\left(\frac{\overline{\varphi}_j}{2}, 1-\frac{\overline{z}_k^4}{\left(4r^2+\overline{z}_k^2\right) \left(r \pm \sqrt{r^2+\overline{z}_k^2}\right)^2} , \frac{4r^2}{4r^2 + \overline{z}_k^2} \right)$ 
\\
 \hline\hline
 $H_{r,z_k}^{r_i,\varphi_j}$ & 
$\cdots$ &
$-\frac{\cos{\theta_M} \left(\sqrt{r^2 + \overline{z}_k^2 } \mp r\right)\left(r\pm \sqrt{r^2 + \overline{z}_k^2 }\right)^2}{r\left|\overline{z}_k\right|\sqrt{r^2 + \overline{z}_k^2 }}$ & 
 $\pm \frac{\cos{\theta_M} \overline{z}_k^4}{r\sqrt{\left(r^2 + \overline{z}_k^2\right)\left(4r^2 + \overline{z}_k^2\right)} \left(r\pm \sqrt{r^2 + \overline{z}_k^2 }\right)}$ 
 \\
 \hline
 $H_{z,z_k}^{r_i,\varphi_j}$ & 
$\cdots$ &
 $\cos{\theta_M}\sign{\overline{z}_k}$ & 
 $0$ 
\end{tabular}
\end{center}
\caption{Required functions and corresponding coefficients for $\mathcal{I} = 214$.}
\label{tab::case214}
\end{table}

\begin{table}
\begin{center}
\begin{tabular}{ c||c|c|c|c|c|c|c } 
& $\sqrt{\overline{r}_i^2 + \overline{z}_k^2}$ & 
$\artanh{\left( \frac{\overline{r}_i}{\sqrt{\overline{r}_i^2 + \overline{z}_k^2}} \right)}$ &
$\artanh{\left( \frac{\overline{z}_k}{\sqrt{\overline{r}_i^2 + \overline{z}_k^2}} \right)}$ &
$E\left(\frac{\overline{\varphi}_j}{2}, -\frac{4rr_i}{\overline{r}_i^2 + \overline{z}_k^2}\right)$ & 
$F\left(\frac{\overline{\varphi}_j}{2}, -\frac{4rr_i}{\overline{r}_i^2 + \overline{z}_k^2}\right)$ &
$\Pi\left( \frac{\overline{\varphi}_j}{2}, -\frac{4rr_i}{\overline{r}_i^2}, -\frac{4rr_i}{\overline{r}_i^2 + \overline{z}_k^2} \right)$ & 
$\cdots$ 
\\
 \hline\hline
 $H_{r,r_i}^{z_k,\varphi_j}$ & 
 $-\frac{\sin\theta_M \sin\overline{\varphi}_M \overline{z}_k}{2 r^2}$ & 
 $0$ 
 & 
 $\frac{\sin\theta_M \sin\overline{\varphi}_M  \left(r^2 - r_i^2\right)}{2r^2}$ & 
 $-\frac{\sin\theta_M \cos\overline{\varphi}_M \overline{z}_k \sqrt{\overline{r}_i^2 + \overline{z}_k^2}}{2 r^2}$ & 
 $\frac{\sin\theta_M \cos\overline{\varphi}_M \overline{z}_k \left(2 r_i^2 + \overline{z}_k^2\right) }{2 r^2 \sqrt{\overline{r}_i^2 + \overline{z}_k^2}}$ & 
 $\frac{\sin\theta_M \cos\overline{\varphi}_M \overline{z}_k \left(r^2 + r_i^2\right) \left(r+r_i\right) }{2 r^2 \overline{r}_i \sqrt{\overline{r}_i^2 + \overline{z}_k^2}}$ & 
 \\ 
 \hline
 $H_{r,\varphi_j}^{r_i,z_k}$ &
 $0$ &
 $0$ &
 $-\sin\theta_M \sin\overline{\varphi}_M$ &
  $0$ &
 $0$ &
  $0$ &
 \\
 \hline
 $H_{r,z_k}^{r_i,\varphi_j}$ &
 $0$ &
 $0$ &
 $0$ &
 $\frac{\cos \theta_M \sqrt{\overline{r}_i^2 + \overline{z}_k^2}}{r}$ &
 $-\frac{\cos \theta_M \left(r^2 + r_i^2 + \overline{z}_k^2\right)}{r \sqrt{\overline{r}_i^2 + \overline{z}_k^2}}$ &
 $0$ &
$\cdots$ 
 \\
\end{tabular}

\begin{tabular}{ c||c|c|c } 
& $\cdots$ & 
$\Pi\left( \frac{\overline{\varphi}_j}{2}, \frac{2r}{r\pm\sqrt{r^2 + \overline{z}_k^2}}, -\frac{4rr_i}{\overline{r}_i^2 + \overline{z}_k^2} \right)$ & 
$\Pi\left( \frac{\overline{\varphi}_j}{2}, 1-\frac{\overline{z}_k^2 \left(\overline{r}_i^2 + \overline{z}_k^2\right)}{\left(\left(r+r_i\right)^2+\overline{z}_k^2\right) \left(r \pm \sqrt{r^2+\overline{z}_k^2}\right)^2} , \frac{4rr_i}{\left(r+r_i\right)^2 + \overline{z}_k^2}  \right)$ 
\\
 \hline\hline
 $H_{r,z_k}^{r_i,\varphi_j}$ &
$\cdots$ & 
 $-\frac{\cos{\theta_M} \left(\sqrt{r^2 + \overline{z}_k^2 } \mp r\right)\left(r_i\pm \sqrt{r^2 + \overline{z}_k^2 }\right)^2}{r\sqrt{\left(r^2 + \overline{z}_k^2\right)\left(\overline{r}_i^2 + \overline{z}_k^2\right) }}$ &
 $\pm \frac{\cos{\theta_M} \overline{z}_k^2 \left(\overline{r}_i^2 + \overline{z}_k^2\right)}{r\sqrt{\left(r^2 + \overline{z}_k^2\right)\left(\left(r + r_i\right)^2 + \overline{z}_k^2\right)} \left(r\pm \sqrt{r^2 + \overline{z}_k^2 }\right)}$
 \\ 
 \hline
 $H_{z,z_k}^{r_i,\varphi_j}$ &
$\cdots$ & 
 $\frac{\cos\theta_M \overline{z}_k \left(r_i \pm \sqrt{r^2 + \overline{z}_k^2}\right)}{\sqrt{\overline{r}_i^2 + \overline{z}_k^2} \left(r \pm \sqrt{r^2 + \overline{z}_k^2}\right)}$ &
 $0$ 
\end{tabular}
\end{center}
\caption{Required functions and corresponding coefficients for $\mathcal{I} = 215$.}
\label{tab::case215}
\end{table}

\begin{table}
\begin{center}
\begin{tabular}{ c||c|c } 
& $\sign{\overline{z}_k}$ & 
$\sign{\overline{z}_k \log \left|\overline{z}_k\right|}$ 
\\
 \hline\hline
 $H_{r,\varphi_j}^{r_i,z_k}$ & 
 $0$ & 
 $-\sin{\theta_M} \sin{\overline{\varphi}_M}$ 
 \\ 
 \hline
 $H_{z,z_k}^{r_i,\varphi_j}$ & 
 $-\cos{\theta_M} \varphi_j$ & 
 $0$
\end{tabular}
\end{center}
\caption{Required functions and corresponding coefficients for $\mathcal{I} = 221$.}
\label{tab::case221}
\end{table}

\begin{table}
\begin{center}
\begin{tabular}{ c||c|c|c|c } 
& $\frac{r_i}{\sqrt{r_i^2 + \overline{z}_k^2}}$ & 
$\artanh{\left( \frac{r_i}{\sqrt{r_i^2 + \overline{z}_k^2}} \right)}$ &
$\frac{\overline{z}_k}{\sqrt{r_i^2 + \overline{z}_k^2}}$ & 
$\artanh{\left( \frac{\overline{z}_k}{\sqrt{r_i^2 + \overline{z}_k^2}} \right)}$ 
\\
 \hline\hline
 $H_{r,r_i}^{z_k,\varphi_j}$ & 
 $0$ & 
 $0$ & 
 $\sin{\theta_M} \left( \frac{\cos{\overline{\varphi}_M}\varphi_j}{2} - \frac{\sin{\overline{\varphi}_M}}{4} \right)$ & 
 $0$ 
 \\
 \hline
 $H_{r,\varphi_j}^{r_i,z_k}$ & 
 $0$ & 
 $0$ & 
 $0$ & 
 $-\sin{\theta_M}\sin{\overline{\varphi}_M}$  
 \\
 \hline
 $H_{\varphi,r_i}^{z_k,\varphi_j}$ & 
 $0$ & 
 $0$ & 
 $\sin{\theta_M} \left( \frac{\sin{\overline{\varphi}_M}\varphi_j}{2} - \frac{\cos{\overline{\varphi}_M}}{4} \right)$ & 
 $0$
 \\
 \hline
 $H_{\varphi,z_k}^{\varphi_j,r_i}$ & 
 $\cos{\theta_M}$ & 
 $-\cos{\theta_M}$ & 
 $0$ & 
 $0$ 
 \\ 
 \hline
 $H_{z,r_i}^{z_k,\varphi_j}$ & 
 $\sin{\theta_M}\sin{\overline{\varphi}_M}$ & 
 $0$ & 
 $0$ & 
 $0$ 
 \\ 
 \hline
 $H_{z,\varphi_j}^{z_k,r_i}$ & 
 $0$ & 
 $-\sin{\theta_M}\sin{\overline{\varphi}_M}$ & 
 $0$ & 
 $0$ 
 \\
 \hline
  $H_{z,z_k}^{r_i,\varphi_j}$ & 
  $0$ & 
  $0$ & 
  $-\cos{\theta_M} \varphi_j$ & 
  $0$
\end{tabular}
\end{center}
\caption{Required functions and corresponding coefficients for $\mathcal{I} = 222$.}
\label{tab::case222}
\end{table}

\begin{table}
\begin{center}
\begin{tabular}{ c||c|c|c|c } 
& $\sign{\overline{z}_k}$ &
$\frac{\sqrt{r^2 + \overline{z}_k^2}}{r}$ & 
$\artanh{\left( \frac{r}{\sqrt{r^2 + \overline{z}_k^2}} \right)}$ & 
$\artanh{\left( \frac{\overline{z}_k}{\sqrt{r^2 + \overline{z}_k^2}} \right)}$ 
\\
 \hline\hline
 $H_{r,\varphi_j}^{r_i,z_k}$ & 
 $0$ & 
 $0$ & 
 $0$ & 
 $-\sin{\theta_M}\sin{\overline{\varphi}_M} $ 
 \\
 \hline
 $H_{\varphi,z_k}^{\varphi_j,r_i}$ &
 $0$ &
 $\cos{\theta_M}$ &
 $-\cos{\theta_M}$ & 
 $0$ 
 \\ 
 \hline
 $H_{z,\varphi_j}^{z_k,r_i}$ & 
  $0$ & 
 $0$ & 
 $-\sin{\theta_M}\sin{\overline{\varphi}_M}$ & 
 $0$
 \\
 \hline
 $H_{z,z_k}^{r_i,\varphi_j}$ &
 $\cos\theta_M \overline{\varphi}_j$ &
 $0$ & 
 $0$ & 
 $0$ 
\end{tabular}
\end{center}
\caption{Required functions and corresponding coefficients for $\mathcal{I} = 223$.}
\label{tab::case223}
\end{table}

\begin{table}
\begin{center}
\begin{tabular}{ c||c|c|c|c|c|c|c } 
& $\sqrt{4r^2 + \overline{z}_k^2}$ & 
$\artanh{\left( \frac{2r}{\sqrt{4r^2 + \overline{z}_k^2}} \right)}$ & 
$\artanh{\left( \frac{\overline{z}_k}{\sqrt{4r^2 + \overline{z}_k^2}} \right)}$ & 
$E\left(\frac{\overline{\varphi}_j}{2}, -\frac{4r^2}{\overline{z}_k^2}\right)$ & 
$F\left(\frac{\overline{\varphi}_j}{2}, -\frac{4r^2}{\overline{z}_k^2}\right)$ & 
$\Pi\left( \frac{\overline{\varphi}_j}{2}, \frac{2r}{r\pm\sqrt{r^2 + \overline{z}_k^2}}, -\frac{4r^2}{\overline{z}_k^2} \right)$ & 
$\cdots$
\\
 \hline\hline
  $H_{r,r_i}^{z_k,\varphi_j}$ & 
  $-\frac{\sin{\theta_M}\sin{\overline{\varphi}_M}\overline{z}_k}{2r^2}$ & 
  $0$ & 
  $0$ & 
  $-\frac{\sin{\theta_M}\cos{\overline{\varphi}_M} \sign{\overline{z}_k}\overline{z}_k^2}{2r^2}$ & 
  $\frac{\sin{\theta_M}\cos{\overline{\varphi}_M} \sign{\overline{z}_k} \left(2r^2 + \overline{z}_k^2\right)}{2r^2}$ & 
  $0$ & 
  \\ 
  \hline
 $H_{r,\varphi_j}^{r_i,z_k}$ & 
 $0$ & 
 $0$ & 
 $-\sin{\theta_M}\sin{\overline{\varphi}_M} $ & 
 $0$ & 
 $0$ & 
 $0$ & 
 \\
 \hline
 $H_{r,z_k}^{r_i,\varphi_j}$ & 
 $0$ & 
 $0$ & 
 $0$ & 
 $\frac{\cos{\theta_M}\left|\overline{z}_k\right|}{r}$ & 
 $-\frac{\cos{\theta_M}\left(2r^2 + \overline{z}_k^2\right)}{r \left|\overline{z}_k\right|}$ & 
 $-\frac{\cos{\theta_M} \left(\sqrt{r^2 + \overline{z}_k^2 } \mp r\right)\left(r\pm \sqrt{r^2 + \overline{z}_k^2 }\right)^2}{r\left|\overline{z}_k\right|\sqrt{r^2 + \overline{z}_k^2 }}$ & 
$\cdots$
 \\
 \hline
  $H_{\varphi,r_i}^{z_k,\varphi_j}$ & 
 $-\frac{\sin{\theta_M}\cos{\overline{\varphi}_M}\overline{z}_k}{2r^2}$ & 
  $0$ & 
  $-\sin{\theta_M} \cos{\overline{\varphi}_M}$ & 
  $\frac{\sin{\theta_M}\sin{\overline{\varphi}_M} \sign{\overline{z}_k}\overline{z}_k^2}{2r^2}$ & 
  $-\frac{\sin{\theta_M}\sin{\overline{\varphi}_M} \sign{\overline{z}_k} \left(4r^2 + \overline{z}_k^2\right)}{2r^2}$ &
  $0$ & 
  \\
\end{tabular}

\begin{tabular}{ c||c|c } 
& $\cdots$ & 
$\Pi\left( \frac{\overline{\varphi}_j}{2}, 1-\frac{\overline{z}_k^4}{\left(4r^2+\overline{z}_k^2\right) \left(r \pm \sqrt{r^2+\overline{z}_k^2}\right)^2} , \frac{4r^2}{4r^2 + \overline{z}_k^2} \right)$ 
\\
 \hline\hline
 $H_{r,z_k}^{r_i,\varphi_j}$ & 
$\cdots$ & 
 $\pm \frac{\cos{\theta_M} \overline{z}_k^4}{r\sqrt{\left(r^2 + \overline{z}_k^2\right)\left(4r^2 + \overline{z}_k^2\right)} \left(r\pm \sqrt{r^2 + \overline{z}_k^2 }\right)}$ 
 \\
\end{tabular}
\end{center}
\caption{Required functions and corresponding coefficients for $\mathcal{I} = 224$.}
\label{tab::case224}
\end{table}

\begin{table}
\begin{center}
\begin{tabular}{ c||c|c|c|c|c|c|c } 
& $\sqrt{(r+r_i)^2 + \overline{z}_k^2}$ & 
$\artanh{\left( \frac{r + r_i}{\sqrt{\left(r + r_i\right)^2 + \overline{z}_k^2}} \right)}$ &
$\artanh{\left( \frac{\overline{z}_k}{\sqrt{\left(r+r_i\right)^2 + \overline{z}_k^2}} \right)}$ &
$E\left(\frac{\overline{\varphi}_j}{2}, -\frac{4rr_i}{\overline{r}_i^2 + \overline{z}_k^2}\right)$ & 
$F\left(\frac{\overline{\varphi}_j}{2}, -\frac{4rr_i}{\overline{r}_i^2 + \overline{z}_k^2}\right)$ &
$\Pi\left( \frac{\overline{\varphi}_j}{2}, -\frac{4rr_i}{\overline{r}_i^2}, -\frac{4rr_i}{\overline{r}_i^2 + \overline{z}_k^2} \right)$ & 
$\cdots$ 
\\
 \hline\hline
 $H_{r,r_i}^{z_k,\varphi_j}$ & 
 $-\frac{\sin\theta_M \sin\overline{\varphi}_M \overline{z}_k}{2 r^2}$ & 
 $0$ 
 & 
 $\frac{\sin\theta_M \sin\overline{\varphi}_M \left(r^2 - r_i^2\right) }{2r^2} $ & 
 $-\frac{\sin\theta_M \cos\overline{\varphi}_M \overline{z}_k \sqrt{\overline{r}_i^2 + \overline{z}_k^2}}{2 r^2}$ & 
 $\frac{\sin\theta_M \cos\overline{\varphi}_M \overline{z}_k \left(2 r_i^2 + \overline{z}_k^2\right) }{2 r^2 \sqrt{\overline{r}_i^2 + \overline{z}_k^2}}$ & 
 $\frac{\sin\theta_M \cos\overline{\varphi}_M \overline{z}_k \left(r^2 + r_i^2\right) \left(r+r_i\right) }{2 r^2 \overline{r}_i \sqrt{\overline{r}_i^2 + \overline{z}_k^2}}$ & 
 \\ 
 \hline
 $H_{r,\varphi_j}^{r_i,z_k}$ &
 $0$ &
 $0$ &
 $-\sin\theta_M \sin\overline{\varphi}_M$ &
  $0$ &
 $0$ &
  $0$ &
 \\
 \hline
 $H_{r,z_k}^{r_i,\varphi_j}$ &
 $0$ &
 $0$ &
 $0$ &
 $\frac{\cos \theta_M \sqrt{\overline{r}_i^2 + \overline{z}_k^2}}{r}$ &
 $-\frac{\cos \theta_M \left(r^2 + r_i^2 + \overline{z}_k^2\right)}{r \sqrt{\overline{r}_i^2 + \overline{z}_k^2}}$ &
 $0$ &
$\cdots$
 \\
\end{tabular}

\begin{tabular}{ c||c|c|c } 
& $\cdots$ &
$\Pi\left( \frac{\overline{\varphi}_j}{2}, \frac{2r}{r\pm\sqrt{r^2 + \overline{z}_k^2}}, -\frac{4rr_i}{\overline{r}_i^2 + \overline{z}_k^2} \right)$ & 
$\Pi\left( \frac{\overline{\varphi}_j}{2}, 1-\frac{\overline{z}_k^2 \left(\overline{r}_i^2 + \overline{z}_k^2\right)}{\left(\left(r+r_i\right)^2+\overline{z}_k^2\right) \left(r \pm \sqrt{r^2+\overline{z}_k^2}\right)^2} , \frac{4rr_i}{\left(r+r_i\right)^2 + \overline{z}_k^2}  \right)$ \\
 \hline\hline
 $H_{r,z_k}^{r_i,\varphi_j}$ &
$\cdots$ &
 $-\frac{\cos{\theta_M} \left(\sqrt{r^2 + \overline{z}_k^2 } \mp r\right)\left(r_i\pm \sqrt{r^2 + \overline{z}_k^2 }\right)^2}{r\sqrt{\left(r^2 + \overline{z}_k^2\right)\left(\overline{r}_i^2 + \overline{z}_k^2\right) }}$ &
 $\pm \frac{\cos{\theta_M} \overline{z}_k^2 \left(\overline{r}_i^2 + \overline{z}_k^2\right)}{r\sqrt{\left(r^2 + \overline{z}_k^2\right)\left(\left(r + r_i\right)^2 + \overline{z}_k^2\right)} \left(r\pm \sqrt{r^2 + \overline{z}_k^2 }\right)}$
 \\ 
 \hline
 $H_{z,z_k}^{r_i,\varphi_j}$ &
$\cdots$ &
 $\frac{\cos\theta_M \overline{z}_k \left(r_i \pm \sqrt{r^2 + \overline{z}_k^2}\right)}{\sqrt{\overline{r}_i^2 + \overline{z}_k^2} \left(r \pm \sqrt{r^2 + \overline{z}_k^2}\right)}$ &
 $0$ 
\end{tabular}
\end{center}
\caption{Required functions and corresponding coefficients for $\mathcal{I} = 225$.}
\label{tab::case225}
\end{table}

\begin{table}
\begin{center}
\begin{tabular}{ c||c|c } 
& $\sign{\overline{z}_k}$ & 
$\sign{\overline{z}_k \log \left|\overline{z}_k\right|}$ 
\\
 \hline\hline
 $H_{r,\varphi_j}^{r_i,z_k}$ & 
 $0$ & 
 $-\sin{\theta_M} \sin{\overline{\varphi}_{Mj}} \cos\overline{\varphi}_j$ 
 \\ 
 \hline
 $H_{\varphi,\varphi_j}^{z_k,r_i}$ &
 $0$ &
  $\sin{\theta_M} \sin{\overline{\varphi}_{Mj}} \sin\overline{\varphi}_j$ 
 \\
 \hline
 $H_{z,z_k}^{r_i,\varphi_j}$ & 
 $-\cos{\theta_M} \varphi_j$ & 
 $0$
\end{tabular}
\end{center}
\caption{Required functions and corresponding coefficients for $\mathcal{I} = 231$.}
\label{tab::case231}
\end{table}

\begin{table}
\begin{center}
\begin{tabular}{ c||c|c|c|c } 
& $\frac{r_i}{\sqrt{r_i^2 + \overline{z}_k^2}}$ & 
$\artanh{\left( \frac{r_i}{\sqrt{r_i^2 + \overline{z}_k^2}} \right)}$ &
$\frac{\overline{z}_k}{\sqrt{r_i^2 + \overline{z}_k^2}}$ & 
$\artanh{\left( \frac{\overline{z}_k}{\sqrt{r_i^2 + \overline{z}_k^2}} \right)}$ 
\\
 \hline\hline
 $H_{r,r_i}^{z_k,\varphi_j}$ & 
 $0$ & 
 $0$ & 
 $\sin{\theta_M} \left( \frac{\cos{\overline{\varphi}_M}\varphi_j}{2} - \frac{\sin{\left(\overline{\varphi}_{Mj} + \overline{\varphi}_j\right)}}{4} \right)$ & 
 $0$ 
 \\
 \hline
 $H_{r,\varphi_j}^{r_i,z_k}$ & 
 $0$ & 
 $0$ & 
 $0$ & 
 $-\sin{\theta_M}\sin{\overline{\varphi}_{Mj}\cos\overline{\varphi}_j}$  
 \\
 \hline
  $H_{r,z_k}^{r_i,\varphi_j}$ &
  $-\cos{\theta_M}\sin\overline{\varphi}_j$  &
 $\cos{\theta_M}\sin\overline{\varphi}_j$ &
  $0$ &
  $0$
  \\
  \hline
 $H_{\varphi,r_i}^{z_k,\varphi_j}$ & 
 $0$ & 
 $0$ & 
 $\sin{\theta_M} \left( \frac{\sin{\overline{\varphi}_M}\varphi_j}{2} - \frac{\cos{\left(\overline{\varphi}_{Mj} + \overline{\varphi}_j\right)}}{4} \right)$ & 
 $0$
 \\
 \hline
  $H_{\varphi,\varphi_j}^{z_k,r_i}$ &
$0$ &
$0$ &
$0$ &
 $\sin{\theta_M}\sin{\overline{\varphi}_{Mj}\sin\overline{\varphi}_j}$ 
  \\
  \hline
 $H_{\varphi,z_k}^{\varphi_j,r_i}$ & 
 $-\cos{\theta_M} \cos\overline{\varphi}_j$ & 
 $\cos{\theta_M} \cos\overline{\varphi}_j$ & 
 $0$ & 
 $0$ 
 \\ 
 \hline
 $H_{z,r_i}^{z_k,\varphi_j}$ & 
 $-\sin{\theta_M}\sin{\overline{\varphi}_{Mj}}$ & 
 $0$ & 
 $0$ & 
 $0$ 
 \\ 
 \hline
 $H_{z,\varphi_j}^{z_k,r_i}$ & 
 $0$ & 
 $\sin{\theta_M}\sin{\overline{\varphi}_{Mj}}$ & 
 $0$ & 
 $0$ 
 \\
 \hline
  $H_{z,z_k}^{r_i,\varphi_j}$ & 
  $0$ & 
  $0$ & 
  $-\cos{\theta_M} \varphi_j$ & 
  $0$
\end{tabular}
\end{center}
\caption{Required functions and corresponding coefficients for $\mathcal{I} = 232$.}
\label{tab::case232}
\end{table}

\begin{table}
\begin{center}
\begin{tabular}{ c||c|c|c|c|c|c|c|c } 
& $1$ &
$\log\left( -r\cos\overline{\varphi}_j + \sqrt{r^2 + \overline{z}_k^2} \right)$ &
$\frac{\sqrt{r^2 + \overline{z}_k^2}}{r}$ & 
$\artanh{\left( \frac{r\cos\overline{\varphi}_j}{\sqrt{r^2 + \overline{z}_k^2}} \right)}$ & 
$\artanh{\left( \frac{\overline{z}_k}{\sqrt{r^2 + \overline{z}_k^2}} \right)}$ & 
$\arctan{\left(\frac{r\sin\overline{\varphi}_j}{\overline{z}_k}\right)}$ &
$\arctan{\left( \frac{\overline{z}_k\cos\overline{\varphi}_j}{\sin\overline{\varphi}_j\sqrt{r^2 + \overline{z}_k^2}} \right)}$ & 
$\cdots$
\\
 \hline\hline
 $H_{r,\varphi_j}^{r_i,z_k}$ & 
 $0$ & 
 $0$ & 
 $0$ & 
 $0$ &
 $-\sin{\theta_M}\sin{\overline{\varphi}_{Mj}}\cos\overline{\varphi}_j $ & 
 $0$ & 
 $\sin{\theta_M}\sin{\overline{\varphi}_{Mj}}\sin\overline{\varphi}_j $ &
 \\
 \hline
 $H_{r,z_k}^{r_i,\varphi_j}$ & 
 $-\cos{\theta_M}\sin\overline{\varphi}_j$ & 
 $\cos{\theta_M}\sin\overline{\varphi}_j$ & 
 $0$ & 
 $0$ & 
 $0$ & 
 $\frac{\cos{\theta_M}\overline{z}_k}{r}$ & 
 $0$ &
$\cdots$
 \\
 \hline
  $H_{\varphi,\varphi_j}^{z_k,r_i}$ &
  $0$ &
  $0$ &
  $0$ &
  $0$ &
  $\sin{\theta_M}\sin{\overline{\varphi}_{Mj}}\sin\overline{\varphi}_j$ &
  $0$ &
  $\sin{\theta_M}\sin{\overline{\varphi}_{Mj}}\cos\overline{\varphi}_j$ &
  \\
  \hline
 $H_{\varphi,z_k}^{\varphi_j,r_i}$ &
 $0$ &
 $0$ &
 $\cos{\theta_M}$ & 
 $-\cos{\theta_M}\cos\overline{\varphi}_j$ &
 $0$ &
 $0$ &
 $0$ &
 \\ 
 \hline
 $H_{z,\varphi_j}^{z_k,r_i}$ & 
 $0$ & 
 $0$ &
 $0$ &
 $-\sin{\theta_M}\sin{\overline{\varphi}_{Mj}}$ & 
 $0$ & 
 $0$ & 
 $0$ & 
 \\
 \hline
 $H_{z,z_k}^{r_i,\varphi_j}$ & 
 $0$ & 
 $0$ & 
 $0$ & 
 $0$ & 
 $0$ & 
 $0$ & 
 $0$ & 
$\cdots$
\end{tabular}

\begin{tabular}{ c||c|c|c } 
& $\cdots$ &
$\Sc\left(2\overline{\varphi}_j, \frac{\sqrt{r^2 + \overline{z}_k^2}}{\left|\overline{z}_k\right|} \right)$ & 
$\Sc\left(\overline{\varphi}_j, \left| \frac{\overline{z}_k}{ r \pm \sqrt{r^2 + \overline{z}_k^2}} \right| \right)$
\\
 \hline\hline
 $H_{r,z_k}^{r_i,\varphi_j}$ & 
$\cdots$ &
 $-\frac{\cos{\theta_M}\overline{z}_k}{r}$ &
 $\frac{\cos{\theta_M}\overline{z}_k}{r}$
 \\
 \hline
 $H_{z,z_k}^{r_i,\varphi_j}$ & 
$\cdots$ &
 $\cos{\theta_M}\sign{\overline{z}_k}$ & 
 $0$ 
\end{tabular}
\end{center}
\caption{Required functions and corresponding coefficients for $\mathcal{I} = 233$.}
\label{tab::case233}
\end{table}

\begin{table}
\begin{center}
\begin{tabular}{ c||c|c|c|c|c|c} 
& $1$ & 
$\sqrt{2r^2 \left(1-\cos\overline{\varphi}_j\right) + \overline{z}_k^2}$ & 
$\log\left( r\left(1-\cos\overline{\varphi}_j\right) + \sqrt{2r^2 \left(1-\cos\overline{\varphi}_j\right) + \overline{z}_k^2} \right)$ &
$\artanh{\left( \frac{r\left(1-\cos\overline{\varphi}_j\right)}{\sqrt{2r^2 \left(1-\cos\overline{\varphi}_j\right) + \overline{z}_k^2}} \right)} $ &
$\artanh{\left( \frac{\overline{z}_k}{\sqrt{2r^2 \left(1-\cos\overline{\varphi}_j\right) + \overline{z}_k^2}} \right)}$ & 
$\cdots$
\\
 \hline\hline
  $H_{r,r_i}^{z_k,\varphi_j}$ & 
  $0$ &
  $-\frac{\sin{\theta_M}\sin{\overline{\varphi}_M}\overline{z}_k}{2r^2}$ & 
  $0$ & 
  $0$ & 
  $0$ & 
  $\cdots$
  \\ 
  \hline
 $H_{r,\varphi_j}^{r_i,z_k}$ & 
 $0$ & 
 $0$ & 
  $0$ & 
 $0$ & 
 $-\sin{\theta_M}\sin{\overline{\varphi}_{Mj}} \cos{\overline{\varphi}_{j}}  $ & 
 $\cdots$
 \\
 \hline
 $H_{r,z_k}^{r_i,\varphi_j}$ & 
 $-\cos{\theta_M} \sin{\overline{\varphi}_{j}}$ &
 $0$ & 
  $\cos{\theta_M} \sin{\overline{\varphi}_{j}}$ &
 $0$ & 
 $0$ & 
 $\cdots$
 \\
 \hline
  $H_{\varphi,r_i}^{z_k,\varphi_j}$ & 
  $0$ &
 $-\frac{\sin{\theta_M}\cos{\overline{\varphi}_M}\overline{z}_k}{2r^2}$ & 
  $0$ & 
  $0$ &
  $-\sin{\theta_M} \cos{\overline{\varphi}_M}$ & 
  $\cdots$
  \\
  \hline
    $H_{\varphi,\varphi_j}^{z_k,r_i}$ &
    $0$ & 
    $0$ &
    $0$ & 
    $0$ &
    $\sin{\theta_M}\sin{\overline{\varphi}_{Mj}} \sin{\overline{\varphi}_j}$ &
   $\cdots$
  \\
  \hline
 $H_{\varphi,z_k}^{\varphi_j,r_i}$ & 
 $0$ & 
 $\frac{\cos{\theta_M}}{r}$ & 
 $0$ &
 $\cos{\theta_M}\cos\overline{\varphi}_j$ & 
 $0$ & 
 \\ 
 \hline
 $H_{z,r_i}^{z_k,\varphi_j}$ & 
 $0$ &
 $\frac{\sin{\theta_M}\sin{\overline{\varphi}_M} }{r}$ & 
 $0$ & 
 $0$ & 
 $0$ & 
 $\cdots$
 \\
\end{tabular}

\begin{tabular}{ c||c|c|c|c|c|c|c } 
& $\cdots$ &
$\arctan{\left(\frac{r\sin\overline{\varphi}_j}{\overline{z}_k}\right)}$ &
$\arctan{\left( \frac{\overline{z}_k\left(1 - \cos\overline{\varphi}_j\right)}{\sin\overline{\varphi}_j\sqrt{2r^2 \left(1-\cos\overline{\varphi}_j\right) + \overline{z}_k^2}} \right)}$ & 
$E\left(\frac{\overline{\varphi}_j}{2}, -\frac{4r^2}{\overline{z}_k^2}\right)$ & 
$F\left(\frac{\overline{\varphi}_j}{2}, -\frac{4r^2}{\overline{z}_k^2}\right)$ & 
$\Pi\left( \frac{\overline{\varphi}_j}{2}, \frac{2r}{r\pm\sqrt{r^2 + \overline{z}_k^2}}, -\frac{4r^2}{\overline{z}_k^2} \right)$ & 
$\cdots$
\\
 \hline\hline
  $H_{r,r_i}^{z_k,\varphi_j}$ & 
$\cdots$ &
  $0$ & 
  $0$ & 
  $-\frac{\sin{\theta_M}\cos{\overline{\varphi}_M} \sign{\overline{z}_k}\overline{z}_k^2}{2r^2}$ & 
  $\frac{\sin{\theta_M}\cos{\overline{\varphi}_M} \sign{\overline{z}_k} \left(2r^2 + \overline{z}_k^2\right)}{2r^2}$ & 
  $0$ & 
  \\ 
  \hline
 $H_{r,\varphi_j}^{r_i,z_k}$ & 
$\cdots$ &
 $0$ & 
 $ -\sin{\theta_M}\sin{\overline{\varphi}_{Mj}} \sin{\overline{\varphi}_{j}}$ &
 $0$ & 
 $0$ & 
 $0$ &
 \\
 \hline
 $H_{r,z_k}^{r_i,\varphi_j}$ & 
 $\cdots$ &
 $\frac{\cos{\theta_M}\overline{z}_k}{r}$ & 
 $0$ & 
  $\frac{\cos{\theta_M}\left|\overline{z}_k\right|}{r }$ & 
 $-\frac{\cos{\theta_M}\left(2r^2 + \overline{z}_k^2\right)}{r \left|\overline{z}_k\right|}$ & 
 $-\frac{\cos{\theta_M} \left(\sqrt{r^2 + \overline{z}_k^2 } \mp r\right)\left(r\pm \sqrt{r^2 + \overline{z}_k^2 }\right)^2}{r\left|\overline{z}_k\right|\sqrt{r^2 + \overline{z}_k^2 }}$ & 
$\cdots$
 \\
 \hline
  $H_{\varphi,r_i}^{z_k,\varphi_j}$ & 
 $\cdots$ &
  $0$ & 
  $0$ &
  $\frac{\sin{\theta_M}\sin{\overline{\varphi}_M} \sign{\overline{z}_k}\overline{z}_k^2}{2r^2}$ & 
  $-\frac{\sin{\theta_M}\sin{\overline{\varphi}_M} \sign{\overline{z}_k} \left(4r^2 + \overline{z}_k^2\right)}{2r^2}$ &
  $0$ & 
  \\
\end{tabular}

\begin{tabular}{ c||c|c } 
& $\cdots$ & 
$\Pi\left( \Sc\left( \overline{\varphi}_j, \frac{\sqrt{4r^2 + \overline{z}_k^2}}{\left|\overline{z}_k\right|} \right), 1-\frac{\overline{z}_k^4}{\left(4r^2+\overline{z}_k^2\right) \left(r \pm \sqrt{r^2+\overline{z}_k^2}\right)^2} , \frac{4r^2}{4r^2 + \overline{z}_k^2}  \right)$ 
\\
 \hline\hline
 $H_{r,z_k}^{r_i,\varphi_j}$ & 
$\cdots$ &
 $\pm \frac{\cos{\theta_M} \overline{z}_k^4}{r\sqrt{\left(r^2 + \overline{z}_k^2\right)\left(4r^2 + \overline{z}_k^2\right)} \left(r\pm \sqrt{r^2 + \overline{z}_k^2 }\right)}$ 
 \\
\end{tabular}
\end{center}
\caption{Required functions and corresponding coefficients for $\mathcal{I} = 234$.}
\label{tab::case234}
\end{table}

\begin{table}
\begin{center}
\begin{tabular}{ c||c|c|c|c|c} 
& $1$ &
$\sqrt{r^2 + r_i^2 - 2 r r_i \cos\overline{\varphi}_j + \overline{z}_k^2}$ & 
$\log\left( r_i - r \cos\overline{\varphi}_j + \sqrt{r^2 + r_i^2 - 2 r r_i \cos\overline{\varphi}_j + \overline{z}_k^2} \right) $ &
$\artanh{\left( \frac{r \cos\overline{\varphi}_j - r_i}{\sqrt{r^2 + r_i^2 - 2 r r_i \cos\overline{\varphi}_j + \overline{z}_k^2}} \right)}$ &
$\cdots$
 \\
 \hline\hline
 $H_{r,r_i}^{z_k,\varphi_j}$ & 
 $0$ &
 $-\frac{\sin\theta_M \sin\overline{\varphi}_M \overline{z}_k}{2 r^2}$ & 
 $0$ & 
 $0$ &
 $\cdots$
 \\ 
 \hline
 $H_{r,\varphi_j}^{r_i,z_k}$ &
 $0$ &
 $0$ &
 $0$ &
 $0$ &
 $\cdots$
 \\
 \hline
 $H_{r,z_k}^{r_i,\varphi_j}$ &
 $-\cos\theta_M\sin\overline{\varphi}_j$ &
 $0$ &
 $\cos\theta_M\sin\overline{\varphi}_j$ &
 $0$ &
 $\cdots$
 \\ 
 \hline
 $H_{\varphi,r_i}^{z_k,\varphi_j}$ &
 $0$ &
 $-\frac{\sin\theta_M \cos\overline{\varphi}_M \overline{z}_k}{2 r^2}$ &
 $0$ &
 $0$ &
$\cdots$
 \\
 \hline
 $H_{\varphi,\varphi_j}^{z_k,r_i}$ &
 $0$ & 
 $0$ &
 $0$ & 
 $0$ & 
 $\cdots$
 \\
 \hline
 $H_{\varphi,z_k}^{\varphi_j,r_i}$ & 
 $0$ &
 $\frac{\cos \theta_M}{r}$ &
 $0$ &
 $-\cos \theta_M \cos\overline{\varphi}_j$ & 
 \\
 \hline
 $H_{z,r_i}^{z_k,\varphi_j}$ & 
 $0$ &
 $\frac{\sin \theta_M \sin \overline{\varphi}_M}{r}$ & 
 $0$ & 
 $0$ &
 $\cdots$
 \\
 \hline
    $H_{z,\varphi_j}^{z_k,r_i}$ & 
    $0$ & 
    $0$ & 
    $0$ &
    $-\sin \theta_M \sin \overline{\varphi}_{Mj}$ & 
 \\
 \hline
 $H_{z,z_k}^{r_i,\varphi_j}$ &
 $0$ &
 $0$ &
 $0$ &
 $0$ &
$\cdots$
\end{tabular}

\begin{tabular}{ c||c|c|c|c|c|c|c } 
& $\cdots$ &
$\artanh{\left( \frac{\overline{z}_k}{\sqrt{r^2 + r_i^2 - 2 r r_i \cos\overline{\varphi}_j + \overline{z}_k^2}} \right)}$ &
$\arctan{\left( \frac{r \sin\overline{\varphi}_j}{\overline{z}_k} \right)}$ &
$\arctan{\left( \frac{\overline{z}_k \left(r\cos\overline{\varphi}_j - r_i\right)}{r \sin\overline{\varphi}_j\sqrt{r^2 + r_i^2 - 2 r r_i \cos\overline{\varphi}_j + \overline{z}_k^2}} \right)}$ &
$E\left(\frac{\overline{\varphi}_j}{2}, -\frac{4rr_i}{\overline{r}_i^2 + \overline{z}_k^2}\right)$ & 
$F\left(\frac{\overline{\varphi}_j}{2}, -\frac{4rr_i}{\overline{r}_i^2 + \overline{z}_k^2}\right)$ &
$\cdots$
\\
 \hline\hline
 $H_{r,r_i}^{z_k,\varphi_j}$ & 
$\cdots$ &
 $\frac{\sin\theta_M \sin\overline{\varphi}_M \left(r^2 - r_i^2\right) }{2r^2} $ & 
  $0$ & 
 $0$ &
 $-\frac{\sin\theta_M \cos\overline{\varphi}_M \overline{z}_k \sqrt{\overline{r}_i^2 + \overline{z}_k^2}}{2 r^2}$ & 
 $\frac{\sin\theta_M \cos\overline{\varphi}_M \overline{z}_k \left(2 r_i^2 + \overline{z}_k^2\right) }{2 r^2 \sqrt{\overline{r}_i^2 + \overline{z}_k^2}}$ & 
$\cdots$
 \\ 
 \hline
 $H_{r,\varphi_j}^{r_i,z_k}$ &
$\cdots$ &
 $-\sin\theta_M \sin\overline{\varphi}_{Mj} \cos\overline{\varphi}_j$ &
 $0$ &
 $\sin\theta_M \sin\overline{\varphi}_{Mj} \sin\overline{\varphi}_j$ &
 $0$ &
 $0$ &
 \\
 \hline
$H_{r,z_k}^{r_i,\varphi_j}$ &
 $\cdots$ &
 $0$ &
 $\frac{\cos \theta_M\overline{z}_k}{r}$ &
 $0$ &
 $\frac{\cos \theta_M \sqrt{\overline{r}_i^2 + \overline{z}_k^2}}{r}$ &
 $-\frac{\cos \theta_M \left(r^2 + r_i^2 + \overline{z}_k^2\right)}{r \sqrt{\overline{r}_i^2 + \overline{z}_k^2}}$ &
$\cdots$
 \\
\end{tabular}

\begin{tabular}{ c||c|c|c|c } 
& $\cdots$ &
$\Pi\left( \frac{\overline{\varphi}_j}{2}, -\frac{4rr_i}{\overline{r}_i^2}, -\frac{4rr_i}{\overline{r}_i^2 + \overline{z}_k^2} \right)$ & 
$\Pi\left( \frac{\overline{\varphi}_j}{2}, \frac{2r}{r\pm\sqrt{r^2 + \overline{z}_k^2}}, -\frac{4rr_i}{\overline{r}_i^2 + \overline{z}_k^2} \right)$ & 
$\Pi\left( \Sc\left(\overline{\varphi}_j, \sqrt{\frac{\left(r+r_i\right)^2 + \overline{z}_k^2}{\overline{r}_i^2 + \overline{z}_k^2}}\right), 1-\frac{\overline{z}_k^2 \left(\overline{r}_i^2 + \overline{z}_k^2\right)}{\left(\left(r+r_i\right)^2+\overline{z}_k^2\right) \left(r \pm \sqrt{r^2+\overline{z}_k^2}\right)^2} , \frac{4rr_i}{\left(r+r_i\right)^2 + \overline{z}_k^2} \right)$ \\
 \hline\hline
 $H_{r,r_i}^{z_k,\varphi_j}$ & 
$\cdots$ &
 $\frac{\sin\theta_M \cos\overline{\varphi}_M \overline{z}_k \left(r^2 + r_i^2\right) \left(r+r_i\right) }{2 r^2 \overline{r}_i \sqrt{\overline{r}_i^2 + \overline{z}_k^2}}$ & 
 $0$ & 
 $0$ 
 \\ 
 \hline
 $H_{r,z_k}^{r_i,\varphi_j}$ &
$\cdots$ &
 $0$ &
 $-\frac{\cos{\theta_M} \left(\sqrt{r^2 + \overline{z}_k^2 } \mp r\right)\left(r_i\pm \sqrt{r^2 + \overline{z}_k^2 }\right)^2}{r\sqrt{\left(r^2 + \overline{z}_k^2\right)\left(\overline{r}_i^2 + \overline{z}_k^2\right) }}$ &
 $\pm \frac{\cos{\theta_M} \overline{z}_k^2 \left(\overline{r}_i^2 + \overline{z}_k^2\right)}{r\sqrt{\left(r^2 + \overline{z}_k^2\right)\left(\left(r + r_i\right)^2 + \overline{z}_k^2\right)} \left(r\pm \sqrt{r^2 + \overline{z}_k^2 }\right)}$
 \\ 
 \hline
 $H_{\varphi,r_i}^{z_k,\varphi_j}$ &
$\cdots$ &
  $\frac{\sin\theta_M \sin\overline{\varphi}_M \overline{z}_k \left(r+r_i\right)^2 }{2 r^2 \sqrt{\overline{r}_i^2 + \overline{z}_k^2}}$ &
    $0$ &
    $0$
 \\
\end{tabular}
\end{center}
\caption{Required functions and corresponding coefficients for $\mathcal{I} = 235$.}
\label{tab::case235}
\end{table}

 \end{landscape}
 
 \section{Supplementary data}
 We provide a Python3 implementation of the formulas derived in the paper for calculating the magnetic field of a cylinder tile.





\begin{thebibliography}{1}



\bibitem{strnat1990}
K. J. Strnat, Proceedings of the IEEE 78 (1990) 923-946.

\bibitem{fastenau1996}
R. H. J. Fastenau, E. J. van Loenen, Journal of Magnetism and Magnetic Materials 157–158 (1996) 1-6.

\bibitem{stekly2000}
Z. J. J. Stekly, IEEE Transactions on Applied Superconductivity 10 (2000) 873-878.

\bibitem{coey2002}
J. M. D.Coey, Journal of Magnetism and Magnetic Materials 248 (2002) 441-456.

\bibitem{mccallum2014}
R. W. McCallum, L.H. Lewis, R. Skomski, M.J. Kramer, I.E. Anderson
Annual Review of Materials Research 44 (2014) 451-477.

\bibitem{hirosawa2017}
S. Hirosawa1, M. Nishino1, S. Miyashita, Adv. Nat. Sci: Nanosci. Nanotechnol. 8 (2017) 013002.

\bibitem{coey2020}
J. M. D.Coey, Engineering 6 (2020) 119-131.

\bibitem{b52}
P. Malagò, F. Slanovc, S. Herzog, S. Lumetti, T. Schaden, A. Pellegrinetti, M. Moridi, C. Abert, D. Suess, M. Ortner, Sensors 20 (2020) 6873.



\bibitem{b31}
C. Treutler, Sensors Actuators A 91 (2001) 2–6.

\bibitem{b32}
U. Ausserlechner, A. Satz, F. Gastinger, US Patent (2014) 8,717,010.

\bibitem{b33}
J. Camacho, V. Sosa, Rev Mexicana Fís E 59 (2013) 8–17.


\bibitem{b25}
R. I. Joseph, E. Schlömann, Journal of Applied Physics 36 (1965) 1579.

\bibitem{b26}
A. Smith, K. K. Nielsen,  D. V. Christensen, C. R. H. Bahl, R. Bjørk, J. Hattel, Journal of Applied Physics 107 (2010) 103910.

\bibitem{b27}
J. Prat-Camps, C. Navau, A. Sanchez, and D.-X. Chen, Journal of Applied Physics 107 (2010) 103910.

\bibitem{b51}
A. Caciagli, R. J. Baars, A. P. Philipse, B. W. M. Kuipers, Journal of Magnetism and Magnetic Materials 456 (2018) 423–432.

\bibitem{b35}
N. Derby, S. Olbert, American Journal of Physics 78 (2010) 229.

\bibitem{b36}
E. Furlani, S. Reznik, A. Kroll, IEEE Trans Magn 31 (1995) 844–51.

\bibitem{b37}
R. Ravaud, G. Lemarquand, S. Babic, V. Lemarquand, C. Akyel, IEEE Transactions on Magnetics 46 (2010) 3585–3590.

\bibitem{b38}
J. T. Conway, IEEE Transactions on Magnetics 46 (2009) 75–81.

\bibitem{b39}
J. Selvaggi, S. Salon, M. Chari, Progress in electromagnetics research (2010) 244–51.

\bibitem{b30}
M. Ortner, L. G. C. Bandeira, SoftwareX 11 (2020) 100466.


\bibitem{b77}
J. W. Jansen, J. P. C. Smeets, T. T. Overboom, J. M. M. Rovers, E. A. Lomonova, IEEE Transactions on Magnetics 50 (2014) 1-7.


\bibitem{b46}
M. Ortner, 9th International Conference on Sensing Technology (ICST 2015) 359-364.

\bibitem{b47}
M. Ortner,  M. Ribeiro , D.  Spitzer,  IEEE Transactions on Magnetics 55 (2019) 1-4.

\bibitem{b49}
M. Galea, L. Papini, H. Zhang, C. Gerada, T. Hamiti, IEEE Transactions on Magnetics 51 (2015) 1-9.


\bibitem{b60}
Z. J. Yang, T. H. Johansen, H. Bratsberg, G. Helgesen, A. T. Skjeltorp, Superconductor Science and Technology 3 (1990) 591-597.


\bibitem{engelherbert2005}
 R. Engel-Herbert, T. Hesjedal, Journal of Applied Physics 97 (2005) 074504. 


\bibitem{b41}
O. Chadebec, J.-L. Coulomb, F. Janet, IEEE Transactions on Magnetics 42 (2006) 515-520.

\bibitem{b53}
M. J. Donahue, D. G. Porter, OOMMF user’s guide, version 1.0. interagency report, NISTIR 6376, national institute of standards and technology, gaithersburg, MD (1999).

\bibitem{b54}
C. Abert. magnum.fd—A finite-difference/fft package for the solution of dynamical micromagnetic problems. https:// github. com/ micro magne tics/ magnum. fd. (2013).

\bibitem{b55}
 P. Heistracher, F. Bruckner, C. Abert, C. Vogler, D. Suess, Journal of Magnetism and Magnetic Materials 503 (2020) 166592.
 
 \bibitem{b56}
 C. Abert, F. Bruckner, C. Vogler, R. Windl, R. Thanhoffer, D. Suess, Journal of Magnetism and Magnetic Materials 387 (2015) 13–18. 
 
 \bibitem{b57}
 C. Abert, L. Exl, F. Bruckner, A. Drews, D. Suess, Journal of Magnetism and Magnetic Materials 345 (2013) 29-35.

\bibitem{b58}
A. Vansteenkiste, B. Van de Wiele, Journal of Magnetism and Magnetic Materials 323 (2011) 2585-2591.

\bibitem{b76}
R. Bjørk, E. B. Poulsen, K. K. Nielsen, A. R. Insinga, Journal of Magnetism and Magnetic Materials 535 (2021) 168057.


\bibitem{b59}
C. Huber, C. Abert, F. Bruckner, M. Groenefeld, S. Schuschnigg, I. Teliban, C. Vogler, G. Wautischer, R. Windl, D. Suess, Scientific Reports 7 (2017) 9419. 


\bibitem{b61}
N. Dupré, O. Dubrulle, S. Huber, J. W. Burssens, C. Schott, G. Close, Proceedings 2 (2018) 763.

\bibitem{b62}
U. Ausserlechner, Progress In Electromagnetics Research B 38 (2012) 71-105. 

\bibitem{iso}
Requirements for the technical representation of magnetic measurement scales in design drawings, E DIN SPEC 91411:2021-03 (2021).

\bibitem{b9}
R. Ravaud, G. Lemarquand, V. Lemarquand, C. Depollier, IEEE Transactions on Magnetics 44 (2008) 1982-1989.

\bibitem{b10}
R. Ravaud, G. Lemarquand, Progress In Electromagnetics Research B 24 (2010) 17–32.

\bibitem{b11}
R. Ravaud, G. Lemarquand, Progress In Electromagnetics Research 94 (2009) 327–341.

\bibitem{b12}
R. Ravaud, G. Lemarquand,, V. Lemarquand, IEEE Transactions on Magnetics 45 (2009) 2920-2926.

\bibitem{b13}
S. Song, B. Li, W. Qiao, C. Hu, H. Ren, H. Yu, Q. Zhang, M. Q.-H. Meng, G. Xu, IEEE Transactions on Magnetics 50 (2014) 1-11.

\bibitem{b14}
K. K. Nielsen, R. Bjørk, Journal of Magnetism and Magnetic Materials 507 (2020) 166799.

\bibitem{b6}
T. Fukushima, Journal of Computational and Applied Mathematics 236 (2012) 1961-1975.

\bibitem{b7}
T. Fukushima, IEEE 22nd Symposium on Computer Arithmetic (2015).

\bibitem{b63}
T. Fukushima, Numerische Mathematik 116 (2010) 687–719.

\bibitem{b64}
T. Fukushima, Journal of Computational and Applied Mathematics 235 (2011) 4140–4148.

\bibitem{b65}
T. Fukushima, Journal of Computational and Applied Mathematics 236 (2011) 1961–1975.

\bibitem{b66}
T. Fukushima, Celestial Mechanics and Dynamical Astronomy 105 (2009) 305.

\bibitem{b67}
T. Fukushima, H. Ishizaki, Celestial Mechanics and Dynamical Astronomy 59 (1994) 237–251.

\bibitem{b68}
T. Fukushima, Mathematics of computation 80 (2011) 1725-1743.

\bibitem{b69}
T. Fukushima, Celestial Mechanics and Dynamical Astronomy 105 (2009) 245. 

\bibitem{b70}
T. Fukushima, Journal of Computational and Applied Mathematics 282 (2015) 71-76. 

\bibitem{b3}
J. D. Jackson, \textit{Classical Electrodynamics}, Thrid Edition, John Wiley \& Sons, Inc. (1999).

\bibitem{b71}
E. Jones, T. Oliphant, P. Peterson, SciPy: Open source scientific tools for {Python}, 2001--, http://www.scipy.org/.


\bibitem{b72}
Ansys\textsuperscript{\textregistered} Academic Research Mechanical, Release 18.1.


\bibitem{b73}
H. Raich, P. Blümler, Concepts in Magnetic Resonance Part B Magnetic Resonance Engineering 23B (2004) 16-25.

\bibitem{b74}
P. V. Trevizoli, J. A. Lozano, G. F. Peixer, J. R. Barbosa Jr., Journal of Magnetism and Magnetic Materials, Volume 395 (2015) 109-122.

\bibitem{b75}
D. Zhu, S. P. Beeby, J. Tudor, N. R. Harris,  Smart Materials and Structures. 21 (2012) 075020.



\bibitem{mathematica}
Wolfram Research, Inc., Mathematica, Version 10.1, Champaign, IL (2015).

\bibitem{rubi}
A. Rich, P. Scheibe, and N. M. Abbasi, Journal of Open Source Software 3 (2018) 1073.

\bibitem{b2}
I. S. Gradshteyn, I. M. Ryzhik, Table of Integrals, Series, and Products, Seventh Edition, Academic Press (2007).

\bibitem{b5}
R. Bulirsch, Numerische Mathematik 7 (1965) 78-90.


\bibitem{b8}
R. Bulirsch, Numerische Mathematik 13 (1969) 305-- 315.


%


































\end{thebibliography}
%

\bibliographystyle{unsrtnat}

%








\end{document}